\documentclass[preprint,12pt,authoryear]{elsarticle}
\usepackage{amssymb}
\usepackage{amsmath}
\usepackage{pdflscape}
\journal{Planetary and Space Science}

\begin{document}

\begin{frontmatter}

\title{Recent Chemo-morphological Coma Evolution of Comet \\ 67P/Churyumov–Gerasimenko} 

\author[label1]{Brian P. Murphy} 
\ead{brian.murphy@ed.ac.uk}

\author[label1]{Cyrielle Opitom} 

\author[label1]{Colin Snodgrass} 

\author[label1,label2]{Sophie E. Deam} 

\author[label1]{Léa Ferellec} 

\author[label4]{Matthew Knight} 

\author[label1]{Vincent Okoth} 

\author[label5,label6]{Bin Yang} 

\affiliation[label1]{organization={Institute for Astronomy, University of Edinburgh,},
            addressline={Royal Observatory}, 
            city={Edinburgh},
            postcode={EH9 3HJ},
            country={United Kingdom}}

\affiliation[label2]{organization={Space Science and Technology Centre, Curtin University},
            addressline={GPO Box U1987}, 
            city={Perth},
            postcode={6845}, 
            state={WA},
            country={Australia}}

\affiliation[label3]{organization={International Centre for Radio Astronomy Research, Curtin University},
            addressline={GPO Box U1987}, 
            city={Perth},
            postcode={6845}, 
            state={WA},
            country={Australia}}

\affiliation[label4]{organization={Physics Department, United States Naval Academy},
            city={Annapolis},
            postcode={21402}, 
            state={MD},
            country={United States}}

\affiliation[label5]{organization={Facultad di Ingeniería y Ciencias, Universidad Diego Portales},
            addressline={Av. Ejército 441}, 
            city={Santiago},
            country={Chile}}

\affiliation[label6]{organization={Planetary Science Institute},
            addressline={1700 E Fort Lowell Rd STE 106}, 
            city={Tucson},
            postcode={85719}, 
            state={AZ},
            country={United States}}

\begin{abstract}
We present VLT/MUSE observations of comet 67P/Churyumov-Gerasimenko during its 2021 perihelion passage, from which we generated simultaneous maps of dust, [OI], C$_2$, NH$_2$, and CN comae across 12 pre- and post-perihelion epochs. These maps reveal the evolutionary and compositional trends of 67P’s coma and further enrich the context and findings of ESA's Rosetta mission. Dust and gas species displayed distinct structures, where NH$_2$ and CN signals were uniquely associated with known dust fans, raising the question of possible correlation to the dust and contributions of extended sources. Localised fitted NH$_2$ scale lengths were 1.5-1.9$\times$ larger than those fitted for the rest of the coma, which is consistent with an extended source component for northern pre-perihelion emissions. In the southern hemisphere, CN was correlated with a prominent and sharp dust structure, potentially revealing an extended source origin via larger dust particles that preserve the CN parent species, as evidenced by higher spectral slopes in the region. Gas maps depicted two distinct evolutionary regimes: (1) evolving H$_2$O ([OI]$^{1}$D) and C$_2$ emissions driven by nucleus sublimation and subsolar insolation, and (2) stable NH$_2$ and CN emissions associated with seasonal dynamics and possible distributed sources. Dust spectral slope maps revealed spectral slope trends consistent with Rosetta findings, while green/red [OI] ratios generally indicate a coma dominated by H$_2$O.
\end{abstract}

\begin{keyword}
Comets \sep Comet Coma \sep Comet Dust \sep IFU Spectroscopy \sep Ground-based Observations 
\end{keyword}

\end{frontmatter}

\section{Introduction} \label{sec:introduction}

Comets are among the most well-preserved remnants of the early Solar System, and encapsulate the primordial conditions present 4.56 billion years ago. These icy bodies, composed of refractory solids, volatile ices, and organic compounds, can therefore provide crucial insights into the primordial solar nebula and the evolution of the Solar System until present. From 2014-2016, the European Space Agency's Rosetta Orbiter investigated the Jupiter Family Comet (JFC) 67P/Churyumov-Gerasimenko, hereafter referred to as 67P, and sought to unravel this formation history. Rosetta operated in close proximity to the comet (10s-100s of kilometres), and probed the volatile and refractory inventories, bulk composition, diurnal and seasonal evolution, nucleus topography, inner coma chemistry and composition, and inner coma morphologies across roughly a third of 67P's 2015 apparition \citep{Taylor:2017}. The orbiter carried 11 instruments and the Philae lander, which itself carried another 9 independent instruments for surface and subsurface investigation. Most important to this work are the orbiter instruments that pertained to coma composition and dust particle classification. Rosetta detected vast reservoirs of volatile parent species such as H$_{2}$O, CO, CO$_{2}$, NH$_{3}$, C$_{2}$H$_{2}$, C$_{2}$H$_{6}$, CH$_{4}$, ammoniated salts and more, see Table 2 of \cite{Rubin:2019,Altwegg:2020,Poch:2020} and references therein, in proximity to 67P using Alice \citep{Stern:2007}, ROSINA (Rosetta Orbiter Spectrometer for Ion and Neutral Analysis) \citep{Balsiger:2007}, VIRTIS (Visible and Thermal Infrared Thermal Imaging Spectrometer) \citep{Coradini:2007}, and OSIRIS (Optical, Spectroscopic, and Infrared Remote Imaging System) \citep{Keller:2007} instruments. Additional classification of several major dust regimes and their compositions were determined \citep{Güttler:2019} through the GIADA (Grain Impact Analyser and Dust Accumulator) \citep{Colangeli:2007}, MIDAS (Micro-Imaging Dust Analysis System) \citep{Riedler:2007}, and COSIMA (Cometary Secondary Ion Mass Analyser) \citep{Kissel:2007} instruments.

\begin{figure}[t]
        \centering
        \includegraphics[width=\textwidth]{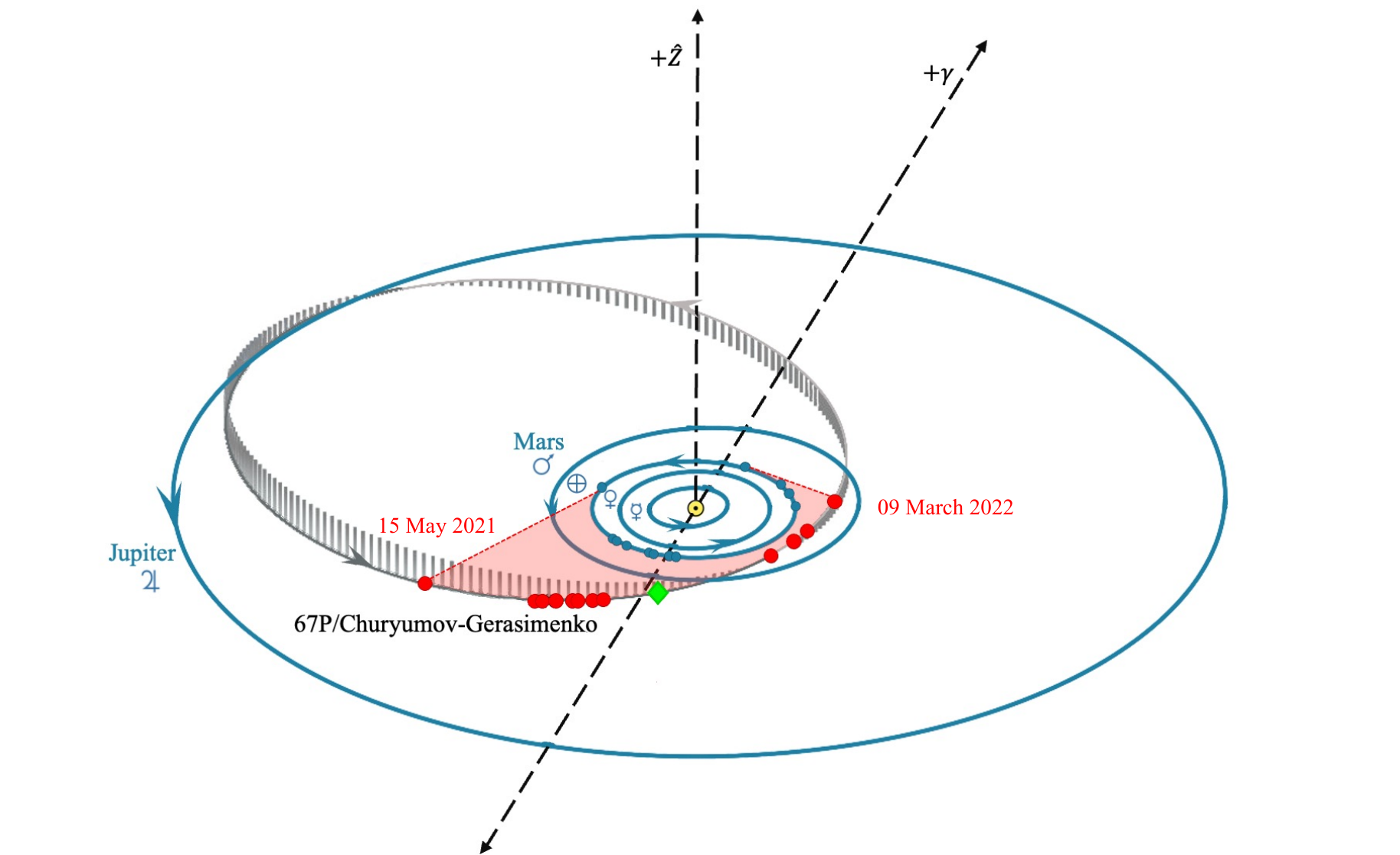}
        \caption{Diagram depicting the orbits of the terrestrial planets, Jupiter, and comet 67P, as of the 2021 apparition. The region shaded red indicates the duration of the observational campaign using MUSE, with the red circles denoting the position of 67P for each observational epoch within the campaign and the blue circles representing the location of Earth for each epoch. The leftmost red circle corresponds to the earliest epoch, with the rightmost corresponding to the latest. The green diamond represents the 2021 perihelion.}\label{fig:orbit}
    \end{figure}

Owing to Rosetta's close proximity to the comet, it was critically important to correlate its findings with those from other ground-based comet studies, which are constrained to observational scales ranging from 10$^3$-10$^5$ kilometres from the nucleus, far outside Rosetta's domain. Extensive ground-based studies leading up to and during the mission were therefore executed to provide high-fidelity ground-truth measurements to better interpret the in-situ results collected by Rosetta \citep{Snodgrass:2017}[and references therein]. In this work, we build upon these previous ground-based studies with 12 follow-up epochs (see Figure \ref{fig:orbit}) of integral field unit (IFU) observations from the Very Large Telescope (VLT) Multi-Unit Spectroscopic Explorer (MUSE) \citep{Bacon:2010} instrument during the most recent 67P apparition in 2021. One key advantage of our analysis is the far better viewing geometry of the 2021 apparition, which resulted in much higher angular resolution (subarcsecond) and closer geocentric distance that allowed for broad chemical species coverage. Therefore, this dataset is uniquely suited as the best comparison to the Rosetta mission, and present our results in the well-established context of post-Rosetta cometary science. 

It is important to note that the species, organics, and salts detected in-situ by Rosetta cannot be directly detected using MUSE or other optical ground-based infrastructure. Despite this, we can observe photo-dissociated daughter products, C$_{2}$, NH$_{2}$, [OI], and CN, that optically fluoresce or emit light in the middle and outer coma, which are a product of parent gases released by the nucleus or coma particles. We therefore underscore the observational disconnect between the deep inner coma that Rosetta probed and the middle and outer coma that we observe, and that linking the two has been a longstanding problem in cometary sciences. 

Various studies have attempted to link the conditions measured by probes on and around comet nuclei to Earth-based coma observations in the optical regime \citep{Crifo:2004,Fulle:2016b}. Although correlating in-situ results to ground based observations is non-trivial for most complex molecules, some photochemical pathways have been identified linking the two domains. Work by \cite{Fink:1991} suggests that NH$_{2}$ radicals are likely products of parent NH$_{3}$ molecules directly sublimated from nucleic ices, however, additional sources could include sublimation of ammoniated salts within dust particles far from the nucleus \citep{DelloRusso:2016,Altwegg:2020,Poch:2020}. Forbidden oxygen is widely accepted to be produced from H$_{2}$O and CO$_{2}$ \citep{Bhardwaj:2002}, however, can also be produced in smaller quantities via CO, O$_{2}$, or any other oxygen-bearing species. The dominant parent species for forbidden oxygen (H$_{2}$O or CO$_{2}$) does vary significantly with heliocentric distance \citep{Raghuram:2014}. Direct relations between HCN (parent) and CN (daughter) have been established by \cite{Combi:1980,Fray:2005}, however, recent work in \cite{Hanni:2020} demonstrate that HCN sublimating directly from the nucleus cannot be responsible for all the CN signal observed by Rosetta, hinting at possible extended sources. Similarly, C$_{2}$ was thought to be derived from the photodissociation of C$_{2}$H$_{2}$, C$_{2}$H$_{6}$, or other higher-order hydrocarbons, however, recent studies have suggested more complex photochemical pathways for its production that are currently not well-defined \citep{Weiler:2012}. Additional possible sources for C$_{2}$ and CN may include large organic molecules, CHON particles, or volatile-bearing dusts.

Finally, Rosetta revealed that the nucleus of 67P is distinctly anisotropic and comprised of two unique hemispheres and two diverse lobes, joined by a volatile-rich neck \citep{Thomas:2015}. \cite{Sierks:2015} showed that the majority of H$_{2}$O activity around perihelion originated from this neck region, designated \textit{Hapi}. However, dispersed activity sources were detected across the northern hemisphere pre-perihelion, and do sustain substantial hyper-volatile sublimation post-perihelion \citep{LeRoy:2015}. Seasonal effects were observed to have one of the greatest influences on surface activity and coma composition, due to the  58$^{\circ}$ rotational obliquity and the rapid onset of an intense southern summer in the 6-months bounding perihelion. This short and highly energetic southern summer has been identified as the likely cause of the pronounced compositional and topographical dichotomy between the northern and southern hemispheres \citep{Filacchione:2022}. The southern summer is also accepted as the driver of global mass transport and resurfacing processes across the northern hemisphere \citep{Keller:2017}, and is responsible for large-scale chemical evolution in the coma, such as the enhanced southern production of CN post-equinox \citep{Opitom:2017}. Understanding these seasonal dynamics is essential for interpreting the evolution of the coma around perihelion.
 
In this study, we present the results derived from our MUSE dataset in the context of the Rosetta mission, focusing on the isolation of simultaneous spectral features within the coma of comet 67P. These include dust particle regimes, the C$_2$ d$^{3}$$\Pi$$_{g}$–a$^{3}$$\Pi$$_{u}$ (0,0) Swan band ($\sim$ 5100~\AA), NH$_2$ $\tilde{A}$$^{2}$A$_{1}$ – $\tilde{X}$$^{2}$B$_{1}$ bands ($\sim$ 6000-7400~\AA), the 5577.339, 6300.304, 6363.776~\AA~ [OI] lines, and CN A$^{2}$$\Pi$–X$^{2}$$\Sigma$$^{+}$ (1,0) Red band ($\sim$9100-9300~\AA) in our datacubes. Utilizing these concurrent molecular maps, we investigated the spatial distribution within the middle and outer coma, identified evolutionary trends pre- and post-perihelion, and uncovered novel relationships and correlations that enhance the understanding of 67P and results of the Rosetta mission.

\section{Observations and Data Processing} \label{sec:observations and data processing}

    \begin{landscape}
    \begin{table}
        \caption{\label{TableObs} Observing Conditions throughout the 67P 2016 and 2021 Apparitions VLT/MUSE Campaigns}
        \begin{centering}
        \begin{tabular}{lcccccccc}
        \hline
        \hline
        Date & Time (UT) & N$_\mathrm{raw}$ & $\alpha$ (deg) & R$_{*}$ (au) & $\Delta$ (au) & Scale (km/$''$) & Airmass & Seeing ($''$) \\
        \hline
        \textbf{2015-Aug-14} &  &  &  &  &  &  &  & \textit{Perihelion} \\
        \hline
        2016-Mar-03 & 03:27-06:08 & 10 & 5.9 & 2.49 & 1.52 & 1107 & 1.4 & 0.9 \\
        2016-Mar-04 & 05:34       & 1 & 5.6 & 2.50 & 1.53 & 1110 & 1.2 & 0.4 \\
        2016-Mar-05 & 04:50-06:29 & 8 & 5.2 & 2.51 & 1.53 & 1110 & 1.3 & 1.1 \\
        2016-Mar-06 & 03:25-04:00 & 4 & 4.9 & 2.52 & 1.54 & 1113 & 1.4 & 1.0 \\
        2016-Mar-07 & 04:39-07:38 & 14 & 4.6 & 2.52 & 1.54 & 1113 & 1.3 & 1.0 \\
        \hline
        2021-May-15 & 07:58-08:32 & 4 & 25.1 & 2.25 & 2.37 & 1719 &  1.8 & 0.8 \\
        2021-Aug-18 & 08:17-08:44 & 3 & 38.9 & 1.52 & 0.85 & 619 & 1.2 & 3.2 \\
        2021-Aug-21 & 06:15-06:58 & 4 & 39.4 & 1.50 & 0.82 & 595 & 1.5 & 0.8 \\
        2021-Aug-31 & 05:52-06:46 & 4 & 40.9 & 1.44 & 0.72 & 521 & 1.6 & 0.8 \\
        2021-Sep-13 & 07:25-08:08 & 4 & 43.2 & 1.36 & 0.61 & 440 & 1.4 & 1.0 \\
        2021-Sep-19 & 07:09-07:52 & 4 & 44.4 & 1.32 & 0.57 & 410 & 1.5 & 0.7 \\
        2021-Sep-27 & 07:41-08:52 & 6 & 46.0 & 1.29 & 0.52 & 375 & 1.4 & 0.9 \\
        2021-Sep-30 & 07:37-08:28 & 4 & 46.6 & 1.28 & 0.50 & 365 & 1.5 & 1.0 \\
        \hline
        \textbf{2021-Nov-11} &  &  &  &  &  &  &  & \textit{Perihelion} \\
        \hline
        2022-Jan-10 & 07:41       & 1 & 13.7 & 1.48 & 0.52 & 376 & 1.9 & 1.7 \\
        2022-Jan-11 & 04:55-05:46 & 4 & 13.0 & 1.48 & 0.52 & 377 & 1.7 & 0.6 \\
        2022-Jan-27 & 02:32-03:31 & 3 & 6.3 & 1.6 & 0.62 & 447 & 1.9 & 0.9 \\
        2022-Feb-06 & 01:59-02:42 & 4 & 10.1 & 1.7 & 0.70 & 509 & 1.8 & 0.6 \\
        2022-Mar-09 & 02:27-03:12 & 4 & 21.9 & 1.9 & 1.10 & 782 & 1.6 & 0.4 \\
        \hline
        \end{tabular}
        \end{centering}
        \textbf{Notes.}{ Date: observation calendar date. Time (UT): time of first and last observations. N$_\mathrm{raw}$: number of exposures. $\alpha$ (deg): average phase angle. R$_{*}$ (au): average heliocentric distance. $\Delta$ (au): average geocentric distance. Scale (km/$''$): sky projected distance/$''$. Airmass: average airmass. Seeing ($''$): average atmospheric seeing. \label{tab:obs}}
    \end{table}
    \end{landscape}

We observed 67P with MUSE on 13 epochs from 15 May 2021 to 09 March 2022. MUSE collected three-dimensional spatio-spectral datacubes, which spanned two spatial (x,y) and one spectral ($\lambda$) dimensions. We also retrieved archival MUSE observations of 67P from 03 to 07 March 2016, observed while the Rosetta mission was still active at 67P. We utilised MUSE in the wide field mode (WFM) without adaptive optics, covering 4750 to 9350~\AA~with an average resolving power of $R\sim $3000 \citep{Bacon:2010}. For 2021 observations of 67P, we exposed MUSE for 600 seconds using non-sidereal tracking, which ensured proper centring and a sufficient signal-to-noise ratio (SNR). We grouped the 600-second exposures into roughly one-hour-long observational blocks, consisting of an Object-Sky-Object-Object-Sky-Object pattern. Occasionally, the observational sequence was interrupted due to poor conditions. The second block of 27 September 2021 was interrupted, so only an additional partial Object-Sky-Object sequence was collected. Similarly, poor conditions interrupted the sequence on 10 January 2022, with only a single Object exposure being recorded. We included the partial sequence from 27 September in our analysis, however, rejected the 10 January observation due to low quality and low signal-to-noise. The sky observations were also exposed for 120 seconds and were positioned 5 to 10 arcminutes away from 67P to limit contamination from the extended coma, and were also coadded into 4-minute-long blocks. To ensure comprehensive coverage, we dithered and rotated by 90$^{\circ}$ counter-clockwise between exposures, which minimised instrumental artefacts. Finally, the observatory staff conducted standard star observations used in the data reduction pipeline, which were exposed for only 120 seconds because of their higher intrinsic flux. The 2016 67P MUSE observations were conducted similarly; however, no rotations were applied between exposures and no sky observations were performed \citep{Opitom:2020}. Conditions for both observational periods can be found in Table \ref{tab:obs}. The 2021-2022 MUSE observations were executed as a filler programme, therefore the observing conditions were suboptimal. For some analyses, these conditions (lunar distance, lunar illumination, thick clouds, etc.) caused significant contamination to the cubes, which reduced signal-to-noise and influenced spectral slope. We do not include these observations in our analysis. 

\subsection{Data Reduction} 
\label{subsec:data reduction}

We reduced the raw datacubes using the standard \texttt{ESO Reflex} automated data reduction pipeline, defined in \cite{Freudling:2013} and \cite{Weilbacher:2020}, and hereafter referred to as Reflex. Reflex's main MUSE data products consist of reduced and calibrated datacubes, white-light images, and pixel tables. The datacubes are the reconstructed 3D (x,y,$\lambda$) MUSE spatio-spectral datacubes. The white light images are datacubes that were collapsed along the $\lambda$ axis, which produced a 2D (x,y) image corresponding to the entire optical regime. The pixel tables are ancillary products that store the intermediate data about the reduction products, and are not used in this analysis. 

Across all datacubes, we utilised Reflex, standard star observations, and other standard calibrations to compute the line spread function, flux calibrations, instrument geometry corrections, illumination corrections, sky subtraction, and initial telluric absorption corrections. The sky subtraction was executed using dedicated sky observations that were reduced and calibrated in tandem with the main datacubes. However, residual sky lines remained in the dataset, but did not cause significant contamination to our sought chemical emission lines and bands, so we did not seek further corrections. The initial Reflex telluric corrections did not provide a satisfactory correction to our datacubes, and did significantly affect the CN A$^{2}$$\Pi$–X$^{2}$$\Sigma$$^{+}$ (1,0) Red band, so we reran reductions without Reflex telluric corrections. We tried a more explicit telluric correction via the \texttt{ESO Molecfit} telluric correction software, outlined in \cite{Kausch:2015} and \cite{Smette:2015}, hereafter referred to as Molecfit. We corrected for telluric contamination in each datacube via our novel method described more in subsection \ref{subsec:datacube processing}. We achieved a much higher quality telluric correction with this method, which resulted in as much as a 25$\times$ increase of CN gas signal in the redder wavelengths. We used this workflow to produce our primary set of science datacubes. 

To investigate the presence of forbidden oxygen [OI] emission in the coma, we reran the Reflex reductions without sky subtractions, performed Molecfit corrections, and forced Reflex to output a fourth data product: the reduced and calibrated sky observation datacubes. We used these sky observations to help isolate cometary [OI] emissions from atmospheric [OI] emissions. This method is necessary due to the high levels of noise introduced when Reflex subtracted the strong [OI] sky emissions. Therefore, we needed to perform our own more nuanced subtraction, first outlined in \cite{Opitom:2020} and described further in subsection \ref{subsubsec:GR}. This constituted our secondary science dataset, which preserved the cometary [OI] signal. 

\subsection{Advanced Datacube Processing} \label{subsec:datacube processing}

\begin{figure}[t!]
        \centering
    	\includegraphics[width=\textwidth]{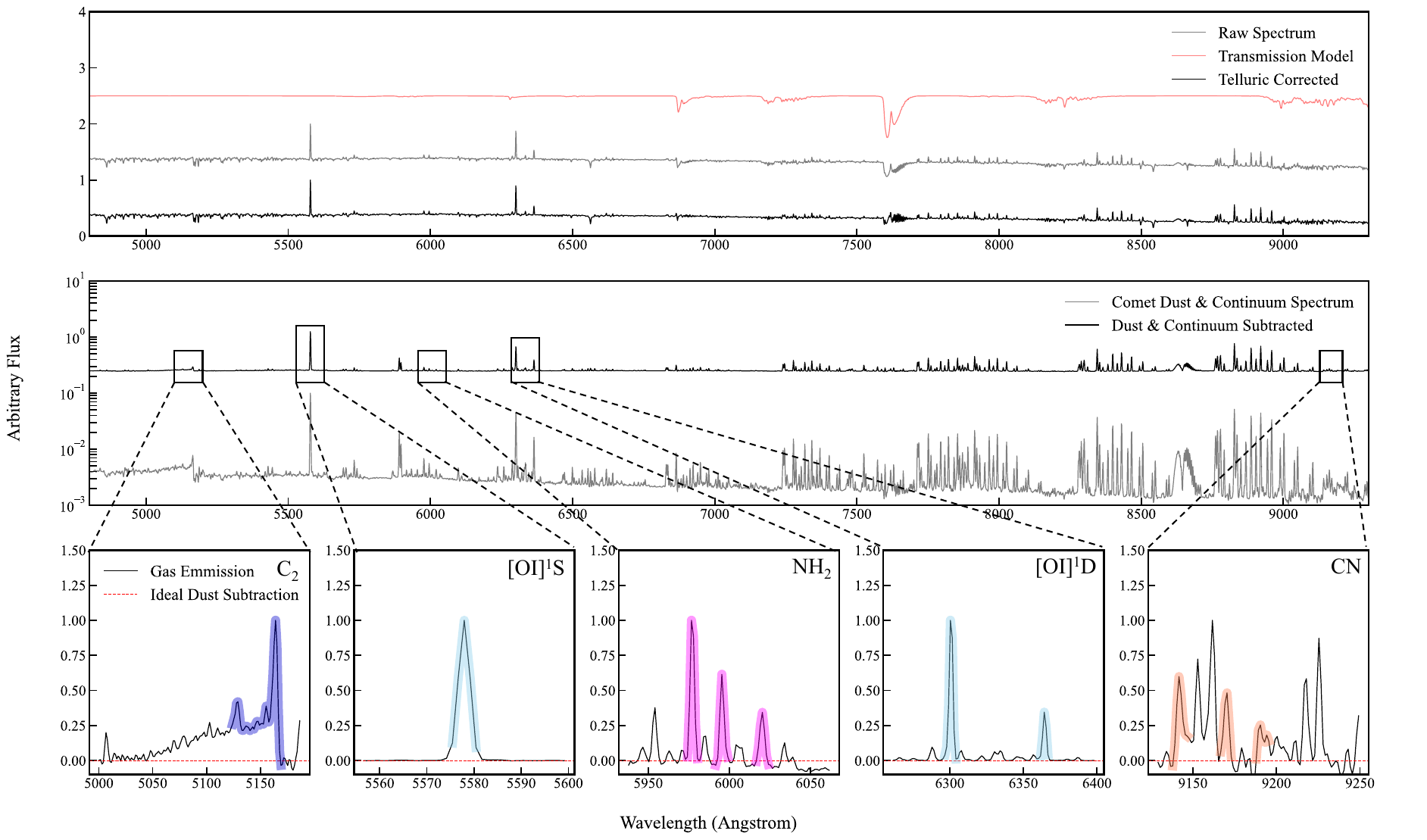}
        \caption{The workflow from initial spectrum to final molecular species isolation and extraction. Top panel: illustration of the telluric correction processing, where the grey line is the raw input spectrum, extracted in $\rho$=10000~km aperture for this figure on 30 September 2021, the red line is the atmospheric transmission model spectrum, and the black line is the final telluric corrected spectrum. Middle panel: continuum isolation and subtraction processing. The grey line is the telluric correct spectrum of 67P from the upper panel. In practice, this process was completed for each of the $\sim$ 96,000 spectra across 48 datacubes, resulting in 4.6 million corrected gaseous spectra. The black line is the final continuum-subtracted spectrum, in which notable gas emission remains. Bottom panels: the residual gas emissions following the dust and solar continuum subtraction, increasing in wavelength from left to right. Red line denotes an ideal subtraction, meaning that if no gas were present the continuum-subtracted spectrum would be the same as this line. Black line is the gas and sky emissions. Highlighted regions correspond to the gas flux summed to create the molecular species maps. The colour of the highlight roughly corresponds to the colour of the species maps, in Figures \ref{fig:OI_enh} through \ref{fig:cn_enh}.} \label{fig:spec_proc}
    \end{figure}

This section describes the post-reduction processing we conducted on each datacube, such as telluric absorption correction, dust and solar continuum corrections and subtraction, molecular species isolation, forbidden oxygen line [OI] extraction, relative spectral slope fitting, and various rebinning and coaddition procedures. Mentioned in subsection \ref{subsec:data reduction}, we leveraged Molecfit to perform corrections of strong telluric absorption features that were heavily contaminating our datacubes, mainly from $\sim$7000 to 9350~\AA. Telluric absorption is caused by the interaction of inbound photons with gaseous molecules, aerosols, and trace gas species in Earth's atmosphere. In our affected spectral region, the primary contaminants are H$_{2}$O and O$_{2}$ absorption bands \citep{Smette:2015}, however, O$_{3}$ Chappuis ozone absorption bands do influence transmittance in the sub-7000~\AA~regime \citep{Chappuis:1880}. We corrected for this by feeding Molecfit an extracted spectrum corresponding to the spaxels within $\rho=10^{4}$ km from the comet optocenter for each datacube. We then executed the Molecfit radiative transfer routines to model synthetic atmospheric transmission spectra of the aforementioned molecular species at each time of observation. The final products consist of a telluric absorption corrected (TAC) spectrum, as well as the stand-alone atmospheric transmission model spectrum. To correct all of the $\sim$96,000 spectra in each datacube, we assumed that the telluric absorption was isotropic \citep{Li:2018,Millan:2024} over the region of sky subtended by MUSE's FoV, and divided each spaxel by the corresponding $\rho=10^{4}$ km atmospheric transmission model. If we consider the bulk contribution of H$_{2}$O and O$_{2}$ telluric absorption occurs within one atmospheric scale length of Earth (8.5 km in height), we assume isotropy over the physical size of the field of view at the top of this scale height, 4.4$\times$4.4 m. Similarly, the O$_{3}$ contribution peaks at $\sim$25 km, roughly subtending 7.3$\times$7.3 m. Finally, if we assume an entire column of the atmosphere, 100 km high, we assume isotropy over 24.3$\times$24.3 m. Through these assumptions, we achieved clean spectra in the final datacubes, as seen in the top panel Figure \ref{fig:spec_proc}.

The main contribution to each spectrum in our datacubes comes from the dust coma, which dominated the observed flux. The dust is non-radiative in the visible regime, so the bulk dust continuum originates from reflected solar light. This reflected solar light still carries signatures intrinsic to the Sun, such as a broad blackbody peak around 6000~\AA~and solar absorption features like the well-documented Fraunhofer lines \citep{Fraunhofer:1817, Kirchhoff:1861}. To correct for this dust and solar continuum, and isolate the gas emission lines and bands, we implement an updated correction workflow that was first outlined in \citep{Opitom:2020}. Therein, \cite{Opitom:2020} analysed MUSE data from the 2016 67P apparition, and attempted to extract gas emissions from the datacubes. The observations were conducted at a heliocentric distance of 2.5~au, and no notable gas emissions were detected above the noise level present, so the spectrum is solely representative of the dust coma and solar reflectance. We leveraged this 2016 reference dust spectrum to isolate the gas emissions in the updated dust and solar continuum subtraction process, outlined below.

We fit a 2D Gaussian to the central 51 pixels of each datacube to locate the optocentre and calculate the centroid of each exposure of the comet. We used the centroid to shift the comet optocentre to the middle of the datacube, and cropped the datacube to 280$\times$280 spaxels, which removed dead spaxels and uneven fringes from each exposure. We then interpolated the 2016 reference dust spectrum to the same sampling and wavelength of the input datacube. In order to model the underlying shape difference between the 2016 reference dust spectrum and input spectrum, we masked the 2016 reference dust spectrum and input comet spectrum to 12 regions that corresponded to regions free of gas emission (dominated by dust). This ensured that we isolated the shape of the dust continuum, without gas contamination. We then divided the masked input spectrum by the masked 2016 reference dust spectrum, which approximated the shape difference between the two. Using this masked difference spectrum, we fit an 11th-order polynomial to the 12 dust regions, which allowed us to best approximate the shape of the smooth underlying dust continuum between these dust regions after much trial and error. We used this technique to extract a final difference spectrum across the entire wavelength range. We used this spectrum to correct the shape and slope deviations of the 2016 reference dust spectrum to the input comet spectrum. Finally, we scaled the shape- and slope-corrected 2016 reference dust spectrum to the flux intensity of the input comet spectrum, and subtracted it from the comet spectrum, which resulted in the final spectrum populated by the residual gas emissions (seen in middle panel of Figure \ref{fig:spec_proc}. This process was conducted iteratively for all spectra in a datacube, and across all datacubes across our campaign. With this technique, we created gas emission datacubes from our entire campaign. Similarly, we also created pure dust continuum datacubes from the fitted dust continua, which we then used for the computation of dust spectral slopes, discussed at the end of this section.

This method robustly disentangles the various spectral contributions in the central wavelength domain, however, has limitations at the blue and red extrema of the instrument, where the strongest C$_{2}$ and CN emission bands are. Despite our optimisations, we were still over- and under-subtracting the dust and solar contributions by $\sim$ 1-5$\%$, so we pivoted to a new technique for the C$_{2}$ Swan band and Red CN bands. We created new datacubes that corresponded to 5000-5300~\AA~ (C$_{2}$ Swan), and 9000-9300~\AA~ (Red CN), selected two gas-free regions on either side of the bands, fit a 1D polynomial to extract the shape and slope, subtracted the dust and solar continuum, and extracted the residual C$_{2}$ and CN emissions, separately. This resulted in a near-perfect subtraction, and was also used for the 5577.339, 6300.304, and 6363.776~\AA~ [OI] lines. 

To probe the environment of the dust coma, and the relative size of particles, we sought to extract maps of the spectral slope from each dust continuum datacube (devoid of gas emission). We first rebinned the $\lambda$ axis (or dispersion axis) by 50~\AA~to decrease the impact of localised deviations, cropped the spectrum to 5000-7000~\AA, and fit a 1D polynomial to each spectrum in each datacube. We took the fitted slope value, converted to units of $\%$/1000~\AA, and cross-checked the value with the slope calculated from the extrema of the rebinned spectrum, using the equation (1) below:

\begin{equation}
     S' = \frac{(F_{2}-F_{1})}{(7000-5000)} * \frac{1}{F_{6000}} * 10000
\end{equation}

\noindent where S' is the spectral slope, F$_{2}$ and F$_{1}$ are the median fluxes in a $\pm$ 200~\AA~window at 7000~\AA~and 5000~\AA~respectively, F$_{6000}$ is the median flux in a $\pm$ 200~\AA~window around 6000~\AA, and 10000 is the conversion factor to $\%$/1000~\AA~\citep{Jewitt:1986}. If both slope measurements were consistent, then the fitted slope was saved into the final slope map. If not, the slope was rejected and the pixel was masked.

Finally, to increase our SNR in the gas emission datacubes, we rebinned them by a factor of 3 to 8 spatially to enhance the detections of C$_2$, NH$_2$, and CN before extracting spatial maps, described in subsection \ref{subsec:image processing}. The dust and [OI] datacubes remained at their original sampling. The spectral slope maps were also rebinned by a factor of 3 before extraction.

\subsection{Image and Map Processing} \label{subsec:image processing}

This section describes the image extraction and processing techniques implemented in the original data cubes, gas emission data cubes, and molecular species maps. To isolate each of the molecular species and dust, we collapsed the gas emission datacubes and original datacubes around specific wavelengths, respectively. For the gas emissions, we summed C$_2$ from 5125-5167~\AA, NH$_2$ peaks at 5973-5982~\AA, 5991-6001~\AA, and 6016-6026~\AA, and CN peaks at 9138.16-9147.7~\AA, 9165.18-9174.7~\AA, and 9185.1-9194.76~\AA. Similarly, we summed the 5577.339~\AA~ [OI] line from 5574.9-5581.2~\AA, 6300.304~\AA \newline [OI] line from 6296.5-6305.8~\AA, and the 6363.776~\AA~ [OI] line from 6361.2-6365.7~\AA. Finally, for the dust maps, we summed 7080-7120~\AA, from the original datacube. We chose all of these wavelengths, seen as highlighted regions in the bottom panels of Figure \ref{fig:spec_proc}, because they were not contaminated by sky lines and they offered the best SNR for the resultant maps.

After we assembled all the molecular maps, we compiled them into nightly sequences, comprised of four 10-minute exposures per stack, one stack per night, on average. The night of 27 September 2021 consisted of six 10-minute exposures, instead of four. We rejected maps if they did not have the comet in the FoV, contained significant star trails through the inner coma, or exhibited major instrumental defects. Details regarding the quantity of rejections, and numbers per stack per night, can be found in Table \ref{tab:obs}.

Our dust and molecular species maps yielded strong detections, however, we sought to probe the underlying structure in each of these components to study morphology, sublimation, and evolutionary processes. We accomplished this by applying division by azimuthal median \cite[cf.][]{Samarasinha:2014}. This technique removed the diffuse contribution of the coma by incrementally calculating the mean of an annulus and dividing it from itself in a series of concentric annuli, which left behind the more elusive substructure. We call this process azimuthal enhancement. We generated azimuthal enhanced Cartesian images, unenhanced Polar images, and enhanced Polar images. We performed analyses on each set of images, which consisted of 12 images each and amounted to 180 final science images. 

For dust features identified in the enhanced images, we initially measured position angles of the underlying substructure from our Polar images, where the x-axis is the azimuthal angle and y-axis is the radial extent. We then summed the first several arcseconds of radial extent to achieve a higher SNR, better profile of the dust coma, and retrieved measurements with the least amount of curvature. Finally, we fit Voigt (Gaussian+Lorentzian) models to our summed profiles, and extracted the angular location of the modelled maxima. This technique is applicable for near-Gaussian profiles, however, we often observed dust structure edges could not be modelled with a Voigt fit. Considering this, we utilised two additional techniques that approximated the position angles for these structures: (1) search for changes in slope by taking the gradient of the profile, and (2) find local deviations from the profile with continuous wavelet transform maps (CWT). Traditionally, CWT is used in the time and frequency domain, however, we applied it in the space and scale domain. This allowed us to measure the position angle (space) of any deviations from the summed radial profile that are caused by underlying dust features. We ensure that these deviations are not instrumental artefacts by varying the scale of the CWT from 1-40, where 1 represents ultra-fine features (likely instrumental defects and noise) and 50 represents large scale structures. We interpret that a structure is real if it is present throughout the scale range we probed, and has consistent position angles across those scales (see Figure \ref{fig:CWT} in \ref{appendix: CWT}). We coupled both techniques to test Voigt measurements and derive additional position angles of fan edges and faint non-Gaussian fans in the post-perihelion regime, discussed further on in this section. 

\section{Methodology and Results} \label{sec:methods and results}

We present the morphological and spectroscopic measurements, calculations, and results, as well as the methods used to perform the requisite analyses, in the following sections.

\subsection{Morphology of Dust and Gas} \label{subsec: morph}

Using methods most recently described in \cite{Murphy:2023b} and \cite{Murphy:2023a}, we tracked the evolving dust and gas morphology throughout the entire observational dataset. 

\subsubsection{Dust Morphology} \label{subsubsec: dust morph}

\begin{figure}[hbt!]
    	\includegraphics[width=\textwidth]{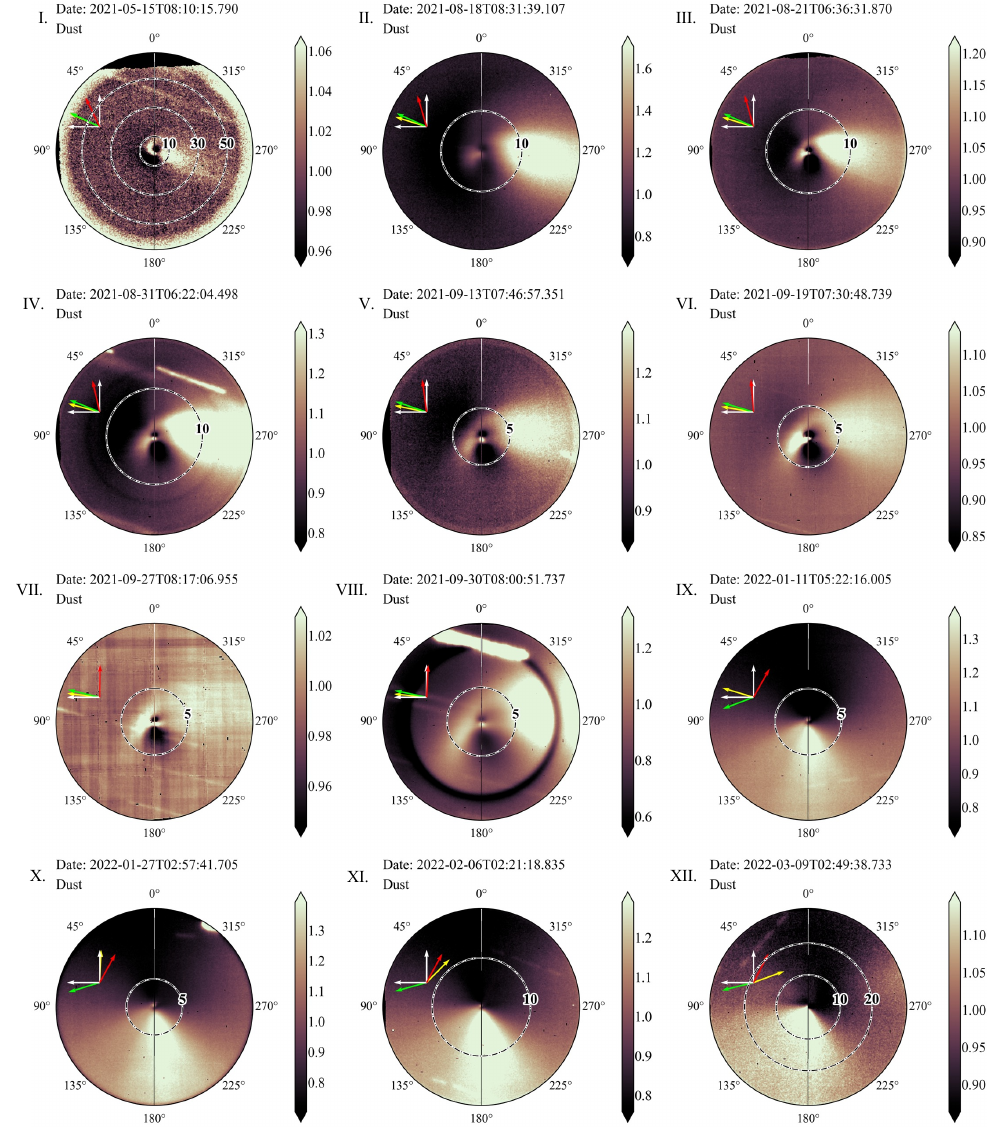}
        \caption{Enhanced dust maps corresponding to the 7080-7120~\AA~ wavelength range, from 15 May 2021 to 09 March 2022. Red arrow denotes sky projection of the Northern pole orientation (2016 Rosetta solution). Lime arrow is the velocity vector. Yellow arrow is the Sun angle. North is up and East is left, denoted by white arrows. Concentric dashed rings denote projected cometocentric distance, $\rho$, as projected onto the sky, in units of 10$^3$ kilometres. No rebinning has been applied.}
        \label{fig:dust_enh}
    \end{figure}

In our first enhanced observations, on 15 May 2021, we found a strong dust signature that extended from a central condensation to around $\rho$=10000 kilometres in the sunward direction (Northeast), seen in panel (I) of Figure \ref{fig:dust_enh}. We denote this structure as the first fan of our observations, Fan A. To the West (anti-sunward direction) a long tail was prominent and extended beyond our FoV of $\rho$=70000 km. The northern profile of the tail was very diffuse with a sharp and distinct southern boundary. The primary axial component of this fan was between the sky-projected northern pole of 67P and sunward angle, around 55$^{\circ}$ NE from reference frame north (RFN). 

    \clearpage

\begin{figure}[hbt!]
    	\includegraphics[width=\textwidth]{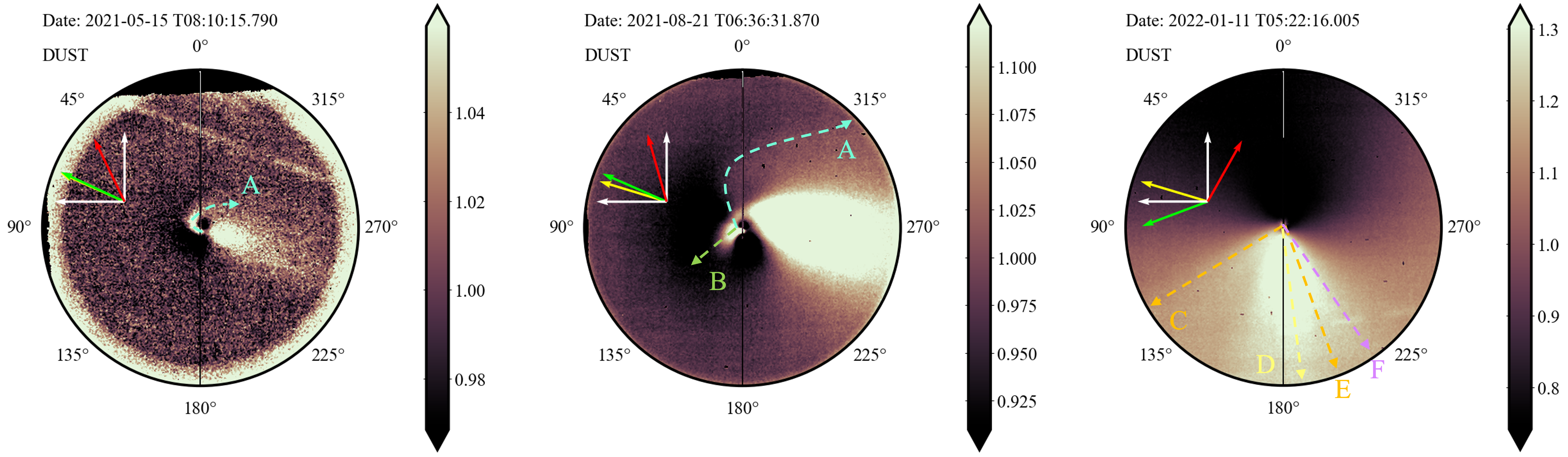}
        \caption{Maps with labelled fans. Left-most panel corresponds to 15 May 2021, where Fan A is present. Center panel corresponds to 21 August 2021, where Fans A and B are present. Right-most panel is 11 January 2022, where Fans C through F are present. Arrows share the same meanings as Figure \ref{fig:dust_enh}.}
        \label{fig:labels}
    \end{figure}

If we assume a spherical projection of the nucleus, we can approximate the latitude range (within $\pm$ $\sim$ 10$^{\circ}$) from which dust and gas structures likely originated. For Fan A, this conversion corresponded to northern latitudes of $\phi\simeq$ 80$^{\circ}$ N. For ease of comparison to the nucleus and other works, we will report both sky-projected and latitudinal position angles for the remainder of this paper. Due to solar radiation pressure (SRP), the dust in Fan A was blown in the anti-sunward direction by $\rho$$\leq$5000 km and overlapped with the northern component of the cometary tail, which possibly contributed the diffuse nature of this structure (see left-most panel in Figure \ref{fig:labels}.) 

By the next epoch of data, 18 August 2021, the viewing geometry was much more favourable ($\Delta = 0.85$ au, geocentric distance). The comet filled the entire FoV, approximately 25000$\times$25000~km. Due to this much closer perspective, further description and characterisation of the tail is limited to direction and relative intensity to adjacent structures. We can only report relative intensities of structures, or contrast, due to both the vastly different observational conditions epoch to epoch and the nature of MUSE as an IFU (difficult to photometrically calibrate). We will focus our remaining characterisation and analysis on dust fans and structures. Fan A persisted after 3 months of inbound evolution, however, the structure appeared to have significantly less contrast to the tail than in the 15 May sighting. Fan A extended linearly to around $\rho$=10000~km, within $\pm$10$^{\circ}$ of the 2015/2016 northern pole solution (still $\phi\simeq$ 80$^{\circ}$ N). Outside of $\rho$=10000~km, the fan curved significantly in the anti-sunward direction and continued to contribute to the diffuse northern component of the tail (see middle panel of Figure \ref{fig:labels}). After 31 August 2021, Fan A was increasingly hard to detect due to the very tight FoV ($\sim$ 10000$\times$10000 km by 30 September) and its diminished apparent brightness and contrast to other dust structures in the FoV. 

During the pre-perihelion period, an additional dust coma structure became evident. A bright and wide dust feature was seen from 18 August onwards. If we assume this structure was comprised of a single active component, it had an opening angle of $\sim$90$^{\circ}$, with a projected position angle of around 90-100$^{\circ}$ from RFN (equatorial latitudes, $\phi\simeq$ 0$^{\circ}$ to -10$^{\circ}$ S). By 21 August, the structure was asymmetrical about the 18 August position measurement, which indicated it is not likely comprised of a single jet or fan, but rather a conglomerate of multiple features projected atop one another. The main component of this structure emanated to the South, hereafter designated Fan B, and extended linearly to beyond $\rho$=5000 km before it gently curved in the anti-sunward direction (see middle panel of Figure \ref{fig:labels}, or panels (II.) and (III.) of Figure \ref{fig:dust_enh}. Fan B was the feature with the most contrast in the dust coma besides the dust tail. From 21 August to 30 September, Fan B increased in contrast, extended to beyond $\rho$=10000 km in radial extent, and remained fixed at a position angle of 140$^{\circ}$$\pm$5$^{\circ}$ from RFN ($\phi\simeq$ -35$^{\circ}$ S). During this period, the tail remained fixed around 270$^{\circ}$ from North, in the anti-sunward direction, and also increased in contrast to the rest of the coma. We note that this increase in contrast and apparent growth of structures is likely due to the decreasing geocentric distance, and thus increasing SNR, rather than actual growth and brightness increases in-system.

    \begin{table}
    \begin{center}
        \caption{\label{tab:jets} Position Angles of Post-perihelion Dust Fans}
        \begin{tabular}{lccc}
        \hline\hline
        Fan & Position Angle ($^{\circ}$) & Latitude $\phi$ ($^{\circ}$, S) &   $\sigma_{PA}$ ($^{\circ}$) \\
        \hline
        \textbf{2022-January-11} &  & \\
        \hline
        Fan-C & 123.1 & -62 & $\substack{+1.6 \\ -1.2}$ \\
        South Pole & 150.7 & -90 & - \\
        Fan-D & 166.3 & -75 & $\substack{+8.3 \\ -7.0}$ \\
        Fan-E & 183.9 & -59 & $\substack{+3.5 \\ -4.6}$ \\
        Fan-F & 210.1 & -33 & $\substack{+1.4 \\ -1.1}$ \\
        \hline
        \textbf{2022-January-27} &  & \\
        \hline
        Fan-C & 128.7 & -68 & $\substack{+2.3 \\ -2.4}$ \\
        South Pole & 151.1 & -90 & - \\
        Fan-D & 164.7 & -76 & $\substack{+5.2 \\ -6.3}$ \\
        Fan-E & 188.0 & -53 & $\substack{+3.6 \\ -4.5}$ \\
        Fan-F & 212.2 & -29 & $\substack{+3.5 \\ -4.6}$ \\
        \hline
        \textbf{2022-February-06} &  & \\
        \hline
        Fan-C & 129.1 & -68  & $\substack{+0.7 \\ -0.6}$ \\
        South Pole & 151.6 & -90 & - \\
        Fan-D & 166.4 & -75 & $\substack{+12.9 \\ -5.7}$ \\
        Fan-E & 187.5 & -54 & $\substack{+5.1 \\ -2.3}$ \\
        Fan-F & 211.8 & -30 & $\substack{+3.6 \\ -2.5}$ \\
        \hline
        \end{tabular}
        \newline
        \newline
    \end{center}
        \textbf{Notes.}{ Fan column refers to the designation of the dust fan, as defined by this work. Position angle was measured from North to East, counter-clockwise, at each of the dates. Nucleus Latitude $\phi$ is the conversion from plane-of-sky position angles and approximate latitudes of nucleus sources, assuming a spherical nucleus. $\sigma_{PA}$ denotes the uncertainty in measurement of the position angles. Dust fans benefit from different upper and lower uncertainties due to their higher SNR signal, which allowed for more robust fits along their radial extent.}
    \end{table}

Post-perihelion, our view of the comet had undergone a change in orbital plane angle from +2$^{\circ}$ to -6$^{\circ}$ and phase angle change from 46$^{\circ}$ to 13$^{\circ}$, which significantly changed the Earth-based perception of the comet dust morphology. From 11 January to 09 March 2022, a large and wide dust structure persisted in the southern half of the dust maps, roughly along the sky-projected South Pole position angle of 67P. No prominent dust features were present in the North throughout this period. The southern dust structure exhibited finer substructure, from which we were able to fit four Voigt models and extract position angles for the first 3 epochs, presented in Table \ref{tab:jets}. We checked our retrieved position angles by computing the CWT of the fitted profile, seen in Figure \ref{fig:CWT} in Appendix section \ref{appendix: CWT}. We could not extract measurements of the substructure by 09 March because the cometary dust tail overlapped with the southern structures and the geocentric distance had increased by a factor of 2, which diminished the quality, contrast, and distinctiveness of the features. From East to West, we fitted four dust structures from the Voigt models, denoted as Fans C, D, E, and F. We found that Fans C and E correlate to structures detected in our CWT maps, however, note that Fans D and F have less precise associations. Considering this is an approximation of source latitude, we suggest that the bulk of post-perihelion dust emission came from between latitudes -10$^{\circ}$ to -30$^{\circ}$ S (Fan F), -40$^{\circ}$ to -50$^{\circ}$ S (Fans C and E), and -60$^{\circ}$ to -80$^{\circ}$ S (Fan D). Intriguingly, Fans C and E were generally symmetric around the South Pole, hinting that they are projections of the same dust structure and activity source, discussed further in Section \ref{sec:discussion}. The position angle of each fan is self-consistent over the period of a month, similar to both pre-perihelion fans, suggesting long-term stability of these structures. However, all post-perihelion fans are linear and devoid of any substantial curvature, unlike the pre-perihelion fans. This is likely due to our viewing geometry, which is nearly head-on (phase angle of $\sim$10$^{\circ}$, so most curvature would likely be away from Earth and imperceivable. Fans C through F are labelled and can be seen in the right-most panel of Figure \ref{fig:labels}, or in the unlabelled panels (IX.) through (XII.) in Figure \ref{fig:dust_enh}.

\subsubsection{Gas Morphology} \label{subsubsec: gas morph}

\begin{figure}[hbt!]
    	\includegraphics[width=\textwidth]{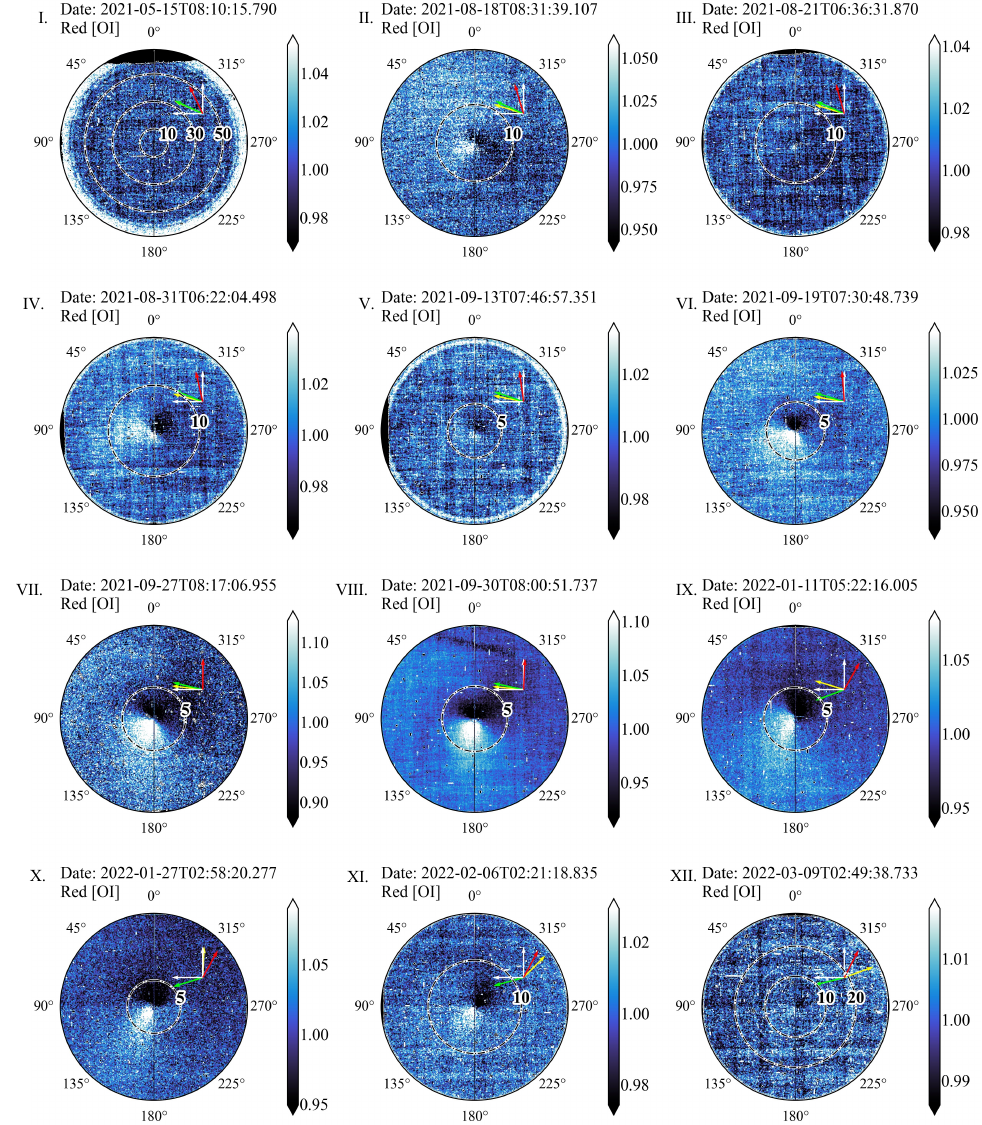}
        \caption{6300.304~\AA~[OI] line from 6296.5-6305.8~\AA, and the 6363.776~\AA~[OI] line from 6361.2-6365.7~\AA~Enhanced red-doublet [OI] maps, from 15 May 2021 to 09 March 2022. Red arrow denotes sky projection of the Northern pole orientation (2016 Rosetta solution). Lime arrow is the velocity vector. Yellow arrow is the Sun angle. North is up and East is left, denoted by white arrows. Concentric dashed rings denote projected cometocentric distance, $\rho$, as projected onto the sky, in units of 10$^3$ kilometres. No rebinning has been applied.}
        \label{fig:OI_enh}
    \end{figure}

\begin{figure}[hbt!]
    	\includegraphics[width=\textwidth]{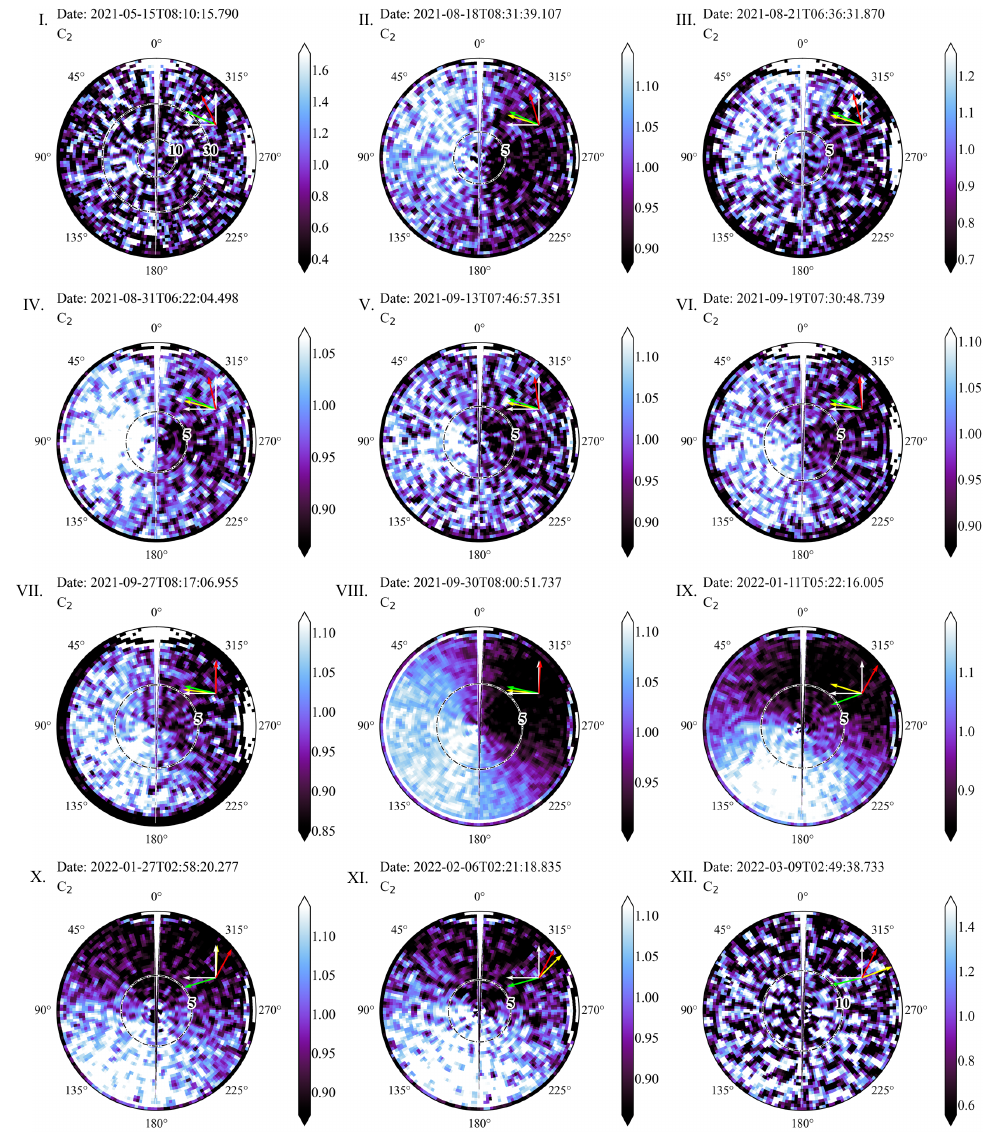}
        \caption{Enhanced C$_2$ maps corresponding to the 5125-5167~\AA~wavelength range, from 15 May 2021 to 09 March 2022. Red arrow denotes sky projection of the Northern pole orientation (2016 Rosetta solution). Lime arrow is the velocity vector. Yellow arrow is the Sun angle. North is up and East is left, denoted by white arrows. Concentric dashed rings denote projected cometocentric distance, $\rho$, as projected onto the sky, in units of 10$^3$ kilometres. Binning is 5$\times$5.} \label{fig:c2_enh}
    \end{figure}

\begin{figure}[hbt!]
    	\includegraphics[width=\textwidth]{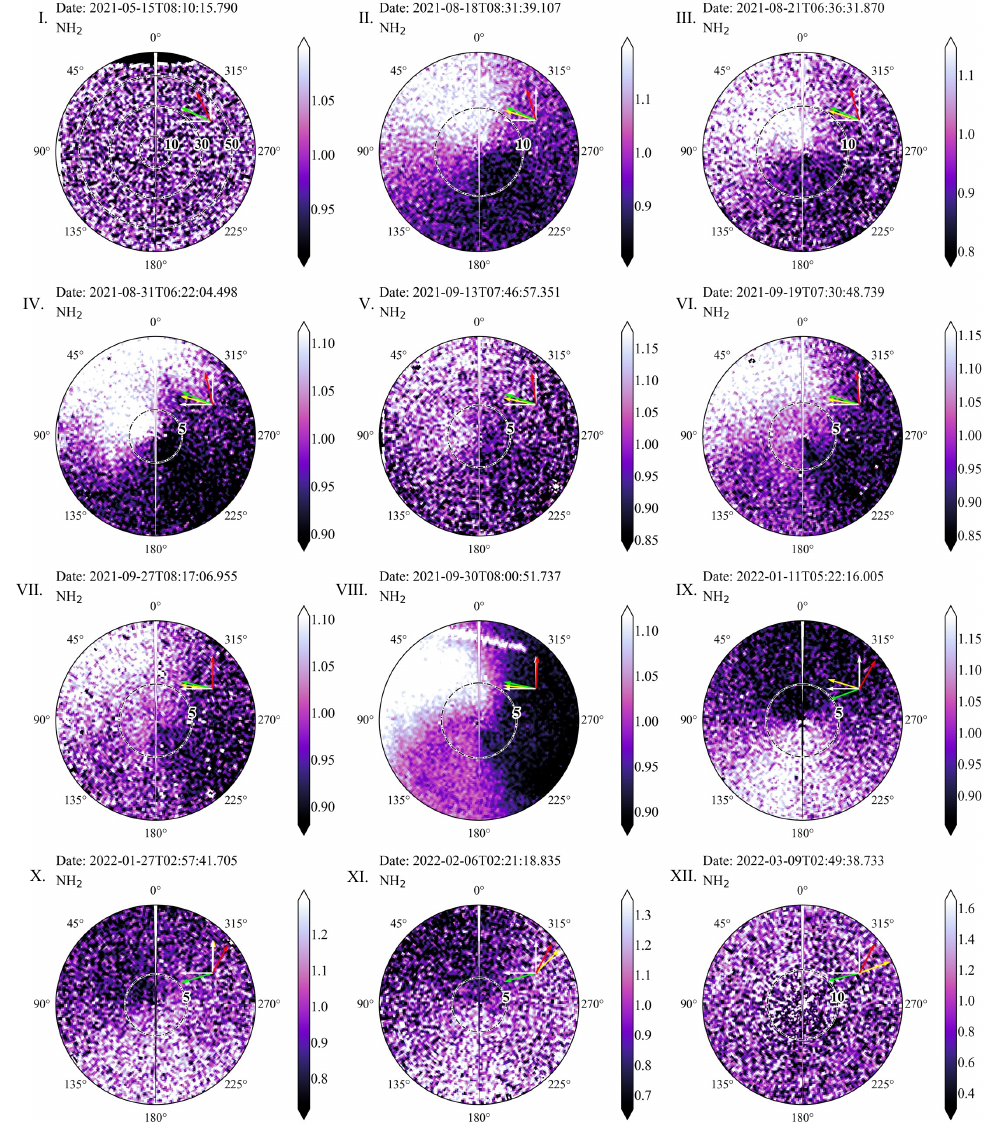}
        \caption{Enhanced NH$_2$ maps corresponding to the 5973-5982~\AA, 5991-6001~\AA, and 6016-6026~\AA~wavelength ranges, from 15 May 2021 to 09 March 2022. Red arrow denotes sky projection of the Northern pole orientation (2016 Rosetta solution). Lime arrow is the velocity vector. Yellow arrow is the Sun angle. North is up and East is left, denoted by white arrows. Concentric dashed rings denote projected cometocentric distance, $\rho$, as projected onto the sky, in units of 10$^3$ kilometres. Binning is 3$\times$3.}
        \label{fig:nh2_enh}
    \end{figure}

\begin{figure}[hbt!]
    	\includegraphics[scale=0.8]{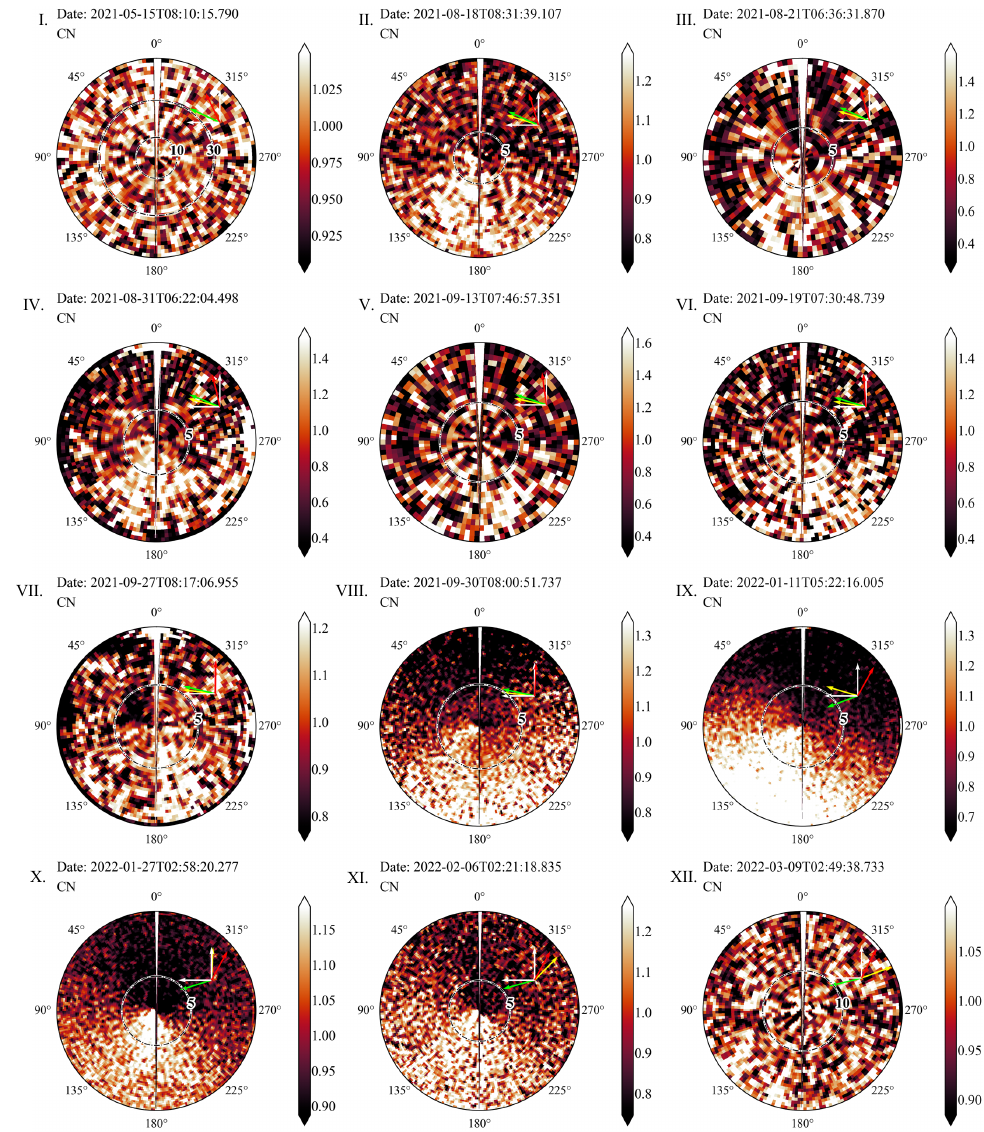}
        \caption{Enhanced CN maps corresponding to the 9138.16-9147.7~\AA, 9165.18-9174.7~\AA, and 9185.1-9194.76~\AA~wavelength ranges, from 15 May 2021 to 09 March 2022. Red arrow denotes sky projection of the Northern pole orientation (2016 Rosetta solution). Lime arrow is the velocity vector. Yellow arrow is the Sun angle. North is up and East is left, denoted by white arrows. Concentric dashed rings denote projected cometocentric distance, $\rho$, as projected onto the sky, in units of 10$^3$ kilometres. Binning ranged from 3$\times$3 to 8$\times$8 depending on the night.}
        \label{fig:cn_enh}
    \end{figure}

Unlike dust Fans A-F, the gas structures exhibit complex and evolving morphologies that differ over species and epoch, as seen in Figures \ref{fig:OI_enh} to \ref{fig:cn_enh}. 

Water sublimation is the primary driver of cometary activity within a few au \citep{Bockelée-Morvan:2004}, and can be traced in the optical regime via green (5577.339~\AA) and red-doublet (6300.304, 6363.776~\AA) electronic transition lines of oxygen atoms from metastable $^{1}$S and $^{1}$D states \citep{Decock:2013}. Destructive excitation of neutral oxygen-bearing molecules, such as H$_{2}$O, CO$_{2}$, and CO, are the primary sources of these forbidden metastable [OI] states \citep{Bhardwaj:2002}. Although CO$_{2}$ and CO dissociation can contribute to both green and red-doublet emissions, at close heliocentric distances the red-doublet flux is dominated by H$_{2}$O dissociation, thus making it ideal to probe the H$_{2}$O morphology of 67P \citep{Decock:2013}. We use the red-doublet lines to construct maps of this morphology over the entire campaign.

    \clearpage

    \clearpage    
    
As seen in Figure \ref{fig:OI_enh}, the first detection of red-doublet [OI] emissions were observed on 18 August 2021, between around 90$^{\circ}$ from RFN (lower northern latitudes, $\phi\simeq$ 25$^{\circ}$ N). This structure was diffuse, localised, and within $\rho$=5000 km from the comet. The next robust detection occurred 13 days later, on 31 August, still centred around 90$^{\circ}$ ($\phi\simeq$ 25$^{\circ}$ N). The structure was still diffuse, prohibiting precise position angle measurements. Despite this, the [OI] structure appeared to increase substantially to around $\rho$=10000 km from the nucleus, which is likely a product of better observing conditions and higher SNR. By 19 September, the [OI] signal had increased and consolidated to the SE quadrant, between 90 and 180$^{\circ}$ (mid-southern latitudes). The southern migration of the [OI] signal persisted over 27 and 30 September, until the signal was nearly aligned with both reference frame South (RFS) and the 67P sky-projected South Pole ($\phi\simeq$ -90$^{\circ}$ S). Post-perihelion, the [OI] structure persisted for the first 3 epochs, from 11 January through 06 February 2022, however, it diminished in relative intensity and radial extent as time progressed. During this time, the [OI] signal was nearly parallel to the 67P sky-projected South pole, around 155$^{\circ}$ from RFN ($\phi\simeq$ -90$^{\circ}$ S), similar to the last pre-perihelion observations. 

Another major molecule studied is the C$_{2}$ radical. After careful dust-subtraction and rebinning procedures, we were able to extract robust C$_{2}$ detections across almost the entire dataset. Similar to [OI], the first detection occurred on 18 August, where a strong preference for the North East quadrant and Eastern hemisphere was observed. From 18 August to 30 September, the C$_{2}$ signal migrated southward, similar to the [OI] signal. By 30 September, the bulk component of C$_{2}$ was centred around 120$^{\circ}$ SE from RFN, and spanned $\pm$80$^{\circ}$ in both directions (mid southern latitudes, $\phi\simeq$ -30 to -40$^{\circ}$ S). Post-perihelion, the C$_{2}$ signal had consolidated to the southern hemisphere and was centred approximately at the South Pole of 67P ($\phi\simeq$ -90$^{\circ}$ S). Both C$_{2}$ and [OI] shared closely aligned and rapidly changing trends over the apparition, thus defining the first major domain of coma evolution; (1) evolving gas structures.

    \clearpage    

    \clearpage    

Unlike C$_{2}$, the NH$_{2}$ radical displayed a relatively stable morphology in the entire campaign. First detected in the North Eastern quadrant of 18 August, the NH$_{2}$ structure was symmetric around the North Pole of 67P (20$^{\circ}$ from RFN), while the southern hemisphere lacked any clear NH$_{2}$ signal after enhancement. From 31 August to 27 September, the NH$_{2}$ structure slowly became asymmetric with respect to 67P's North Pole. The NH$_{2}$ signal had a noticeable counter-clockwise curvature of the main northern structure, and the appearance of a weaker southern component (first seen in (VI.) panel of Figure \ref{fig:nh2_enh}. By 30 September, the highly curved northern NH$_{2}$ structure dominated the field, and the relatively less-intense southern NH$_{2}$ structure displayed no obvious curvature in either direction. Notably, throughout this entire period, the base of the curved northern NH$_{2}$ structure remained fixed at the North Pole of 67P ($\phi\simeq$ 90$^{\circ}$ N), while the nascent southern structure appeared to radiate from near the South Pole of 67P. Similar to both [OI] and C$_{2}$, the post-perihelion morphology of NH$_{2}$ was consolidated to the southern hemisphere of our maps. However, unlike the previous molecules, the NH$_{2}$ signal was spread across the entire hemisphere (opening angles $\leq$230$^{\circ}$), not consolidated into primarily one quadrant, and spread into the NW quadrant by 06 February. 

The CN radical was the final molecule we searched for in our datacubes, and was first detected on 18 August in the southern hemisphere around $\sim$175$^{\circ}$$\pm$20$^{\circ}$ from RFN (South Pole, $\phi\simeq$ -90$^{\circ}$ S). Due to the observational and instrumental constraints discussed in subsection \ref{subsec:datacube processing}, we can only report general trends in CN morphology for most of the campaign, however, we have sufficient SNR to more robustly report observations around perihelion  (30 September 2021 - 06 February 2022) and in the middle of the southern summer, when the CN signal was strongest. For most of the pre-perihelion CN detections, the signal was primarily focused in the southern hemisphere, around the South Pole. Notably, on 30 September, exceptional conditions and proximity to perihelion allowed for a higher SNR CN map to be extracted. Observed within this map is a narrow, curved CN structure centred around 135$^{\circ}$ SE from RFN (mid-southern latitudes, $\phi\simeq$ -45$^{\circ}$ S). Within $\rho$=5000 km, the structure is extremely narrow, linear, and focused, however, it expanded to fill almost the entire southern hemisphere outside of this range. The component outside of $\rho$=5000 km curved toward the anti-sunward direction and dominated the FoV from $\sim$110$^{\circ}$-300$^{\circ}$ from RFN. Surprisingly, this CN structure was remarkably similar to the dust structure designated Fan B. Wary of possible dust contamination, we ran tests to determine if any residual dust had been left in the CN maps post-dust-subtraction, described in Appendix section \ref{appendix: gas-dust}. We found no notable dust residuals from Fan B in the CN maps that correlated to the CN structure described above. Therefore, a strong case can be made for the correlation of Fan B dust and presence of verified CN fluorescence, which could imply that the dust particles in Fan B acted as an extended source for CN production (discussed further in Section \ref{sec:discussion}). Finally, post-perihelion observations depict a symmetric CN structure that was centred around $\sim$165$^{\circ}$ from RFN, congruent with the South Pole of 67P ($\phi\simeq$ -90$^{\circ}$ S). Together with the dust fans, the generally unchanging morphology of NH$_{2}$ and CN constitute the second domain of coma evolution we identify; (2) stable dust and gas structures. We note that our rebinning procedures reduce our spatial resolution for most pre-perihelion nights (except 30 September) to the point that we cannot definitively say that the CN signal does not evolve on small scales within the southern hemisphere, beyond what we can measure, however, we do not find evidence for such evolution at our sensitivity limits.  

\subsection{Dust Environment} \label{subsec: dust}

    \begin{figure}[t!]
    	\includegraphics[width=\textwidth]{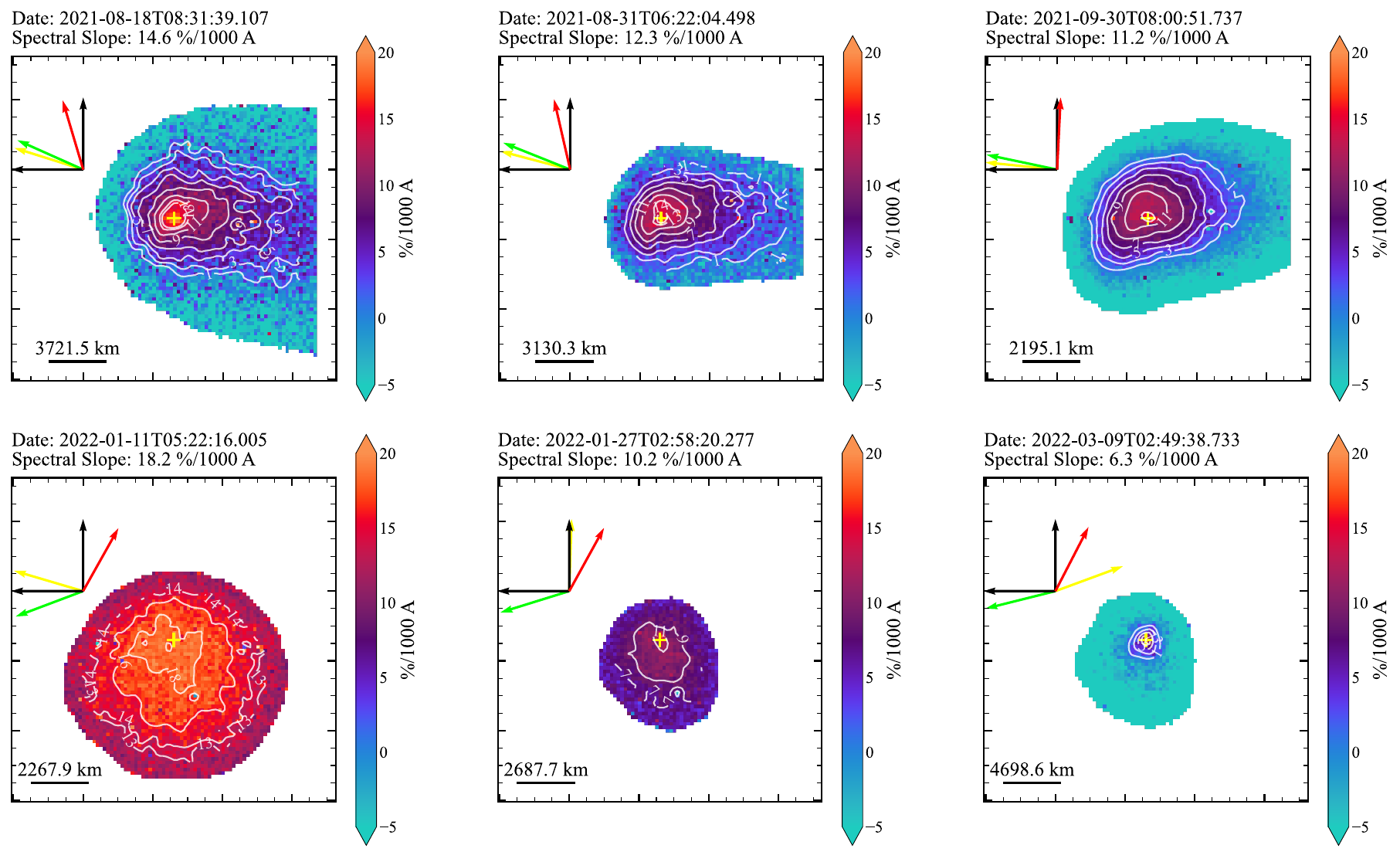}
        \caption{Spectral slope maps from 5000-7000~\AA, normalised 6000~\AA, and expressed in units of $\%$/1000~\AA. Pre-perihelion observations are presented across the top panels, post-perihelion across the bottom panels. These dates were selected based on high signal-to-noise, clear atmospheric conditions, low lunar illumination, and high lunar distance. Spectral slope values on the top axis are given in a 3000~km aperture.}
        \label{fig:slopes}
    \end{figure}

In addition to the dust maps we presented in subsection \ref{subsubsec: dust morph}, we created spectral slope maps to probe the dust particle size distribution of the coma throughout the apparition. Using methods described in subsection \ref{subsec:datacube processing}, we were able to construct three reliable pre-perihelion and two post-perihelion maps. Due to suboptimal conditions (96$\%$ lunar illumination, 4$^{\circ}$ lunar separation, etc.), five of our datacubes had considerable red or blue contamination that made the retrieval of the original spectral slope untenable. Therefore, we only present the maps in which conditions were mostly favourable, and no notable contamination was present. We masked out the background sky through a percentile clip in the original datacubes, in which values less than the 85th percentile were masked, seen in Figure \ref{fig:slopes}. We applied 1$\sigma$ smoothed contours to further highlight the distinct regimes and structures in our spectral slope maps, from 1 to 18$\%$/1000~\AA~pre-perihelion and post-perihelion. Despite the 1$\sigma$ smoothing, some hot pixels and star trails still persist in the contours, and should be ignored. 

Shown in Figure \ref{fig:slopes}, the pre-perihelion maps depict a 6-week period in which the spectral slopes of the coma steadily evolved as 67P approached perihelion and as new regions of the nucleus became active. On 18 August, the dust coma exhibited predominantly red spectral slopes, ranging from 9$\%$/1000~\AA~to 13$\%$/1000~\AA, with a central condensation containing slopes exceeding 15$\%$/1000~\AA. The coma was elongated in the West, corresponding to the anti-sunward tail. A central structure with slopes of 9-11$\%$/1000~\AA~extended approximately 3000 km from the nucleus along this axis. Overall, the coma appeared relatively uniform, with no significant substructure aside from the western elongation. By 31 August, the coma exhibited a 2-3$\%$/1000~\AA~bluer spectral slope globally and the extent of regions with higher spectral slopes had diminished significantly, despite a 1.2$\times$ smaller FoV. While still elongated along the western axis, a new coma structure correlating with the position of Fan B had appeared in the southeastern quadrant. This structure manifested as a ridge in the 11, 9, and 7 $\%$/1000~\AA~contours and extended a few thousand kilometres to the SE. This trend continued into 30 September, where global spectral slopes were considerably lower, the extent of high slope regions ($>$11 $\%$/1000~\AA) were confined to the central thousand kilometres of the coma, and the structure synonymous with Fan B exhibited higher slopes further from the nucleus than surrounding regions, likely suggesting evidence of larger, longer living, or different compositions of dust particles in this region (see discussion in section \ref{sec:discussion}). Despite the significant spectral slope shift in the outer regions of the coma from August to September, the colours within 3000 km maintained relatively high slopes, from 14.6$\pm$2$\%$/1000~\AA~on 18 August, to 12.3$\pm$1$\%$/1000~\AA~on 31 August, and 11.2$\pm$1$\%$/1000~\AA~on 30 September.

As mentioned in subsection \ref{subsubsec: dust morph}, post-perihelion observations are affected by considerable changes in viewing geometry, such as lower phase angles and an orbital plane change. These, in tandem with the orbital position of Earth, change the view geometry of the comet from side-on (pre-perihelion) to nearly head on (post-perihelion). With this context, we report the post-perihelion spectral slope maps in Figure \ref{fig:slopes}. 

On 11 January, the coma exhibited large homogenous regions of high spectral slopes on the order of 14 to 18 $\%$/1000~\AA~in the southern hemisphere, which aligned with the South Pole of 67P. Intriguingly, by 27 January and 06 February the coma slopes had decreased by around 8-10$\%$/1000~\AA~on average. This is a large change across the matter of just a few weeks, and is contrary to the slowly evolving behaviour we saw pre-perihelion and expected post-perihelion. It could be possible that spectral slopes from 11 January are atypical of the post-perihelion trends, and we discuss this in Section \ref{sec:discussion}. Unlike the pre-perihelion trend of bluer spectral slopes closer to the perihelion, the reciprocal was true of post-perihelion measurements, where redder slopes are present in proximity to the Sun while bluer slopes slowly evolve with increasing heliocentric distance and decreasing activity. No major substructure was visible throughout this period, likely due to the viewing geometry, despite the presence of multiple known dust structures (Fans C through F). Spectral slopes within the central 3000 km on 11 January, 27 January, and 06 February were 18.2$\pm$2$\%$/1000~\AA, 10.2$\pm$1$\%$/1000~\AA, and 8.3$\pm$1$\%$/1000~\AA, respectively. 

\subsection{Gas Environment and Species Heritage} \label{subsec: gas}

Mentioned in Section \ref{sec:introduction}, the molecules we observe in our gas maps are not primary products directly sublimated from the nucleus, but rather are secondary, tertiary, or even higher-order products resulting from photochemical, collisional, or thermal degradation processes in the coma. Tracing the formational heritage of these molecules is non-trivial, however, insights can be gained through comparing specific line ratios and employing idealised coma models. In the following section, we present the ratio of green to red-doublet forbidden oxygen [OI] emission lines to dissect the dominance of H$_{2}$O vs CO$_{2}$ in the coma over different heliocentric distances. We also present the application of radial Haser profile fitting to determine possible contribution of extended sources, like icy particles or dust, to the gas morphologies we observe. Finally, we attempted to calculate gas-to-dust ratios of NH$_{2}$ and CN to search for dependence on dust or signs of originating from different extended sources, however, we did not find any notable general trends pre- and post-perihelion.

\subsubsection{G/R Ratio} \label{subsubsec:GR}

The ratio of green to red-doublet emission intensities can approximate if the coma is dominated by H$_{2}$O, or CO$_{2}$ and CO \citep{Festou:1981, Decock:2013, Raghuram:2014}, so long as the FoV is larger than the collisional area of the coma. We measured the intensities and calculated the ratio through a three-step process: (1) fit Gaussians to the green and red-doublets of both dust-subtracted 67P and sky-observation datacubes, (2) isolate 67P [OI] emissions by subtracting the sky [OI] emissions, and (3) compute the ratios and uncertainties. 

\begin{figure}[t!]
        \centering
    	\includegraphics[scale=0.38]{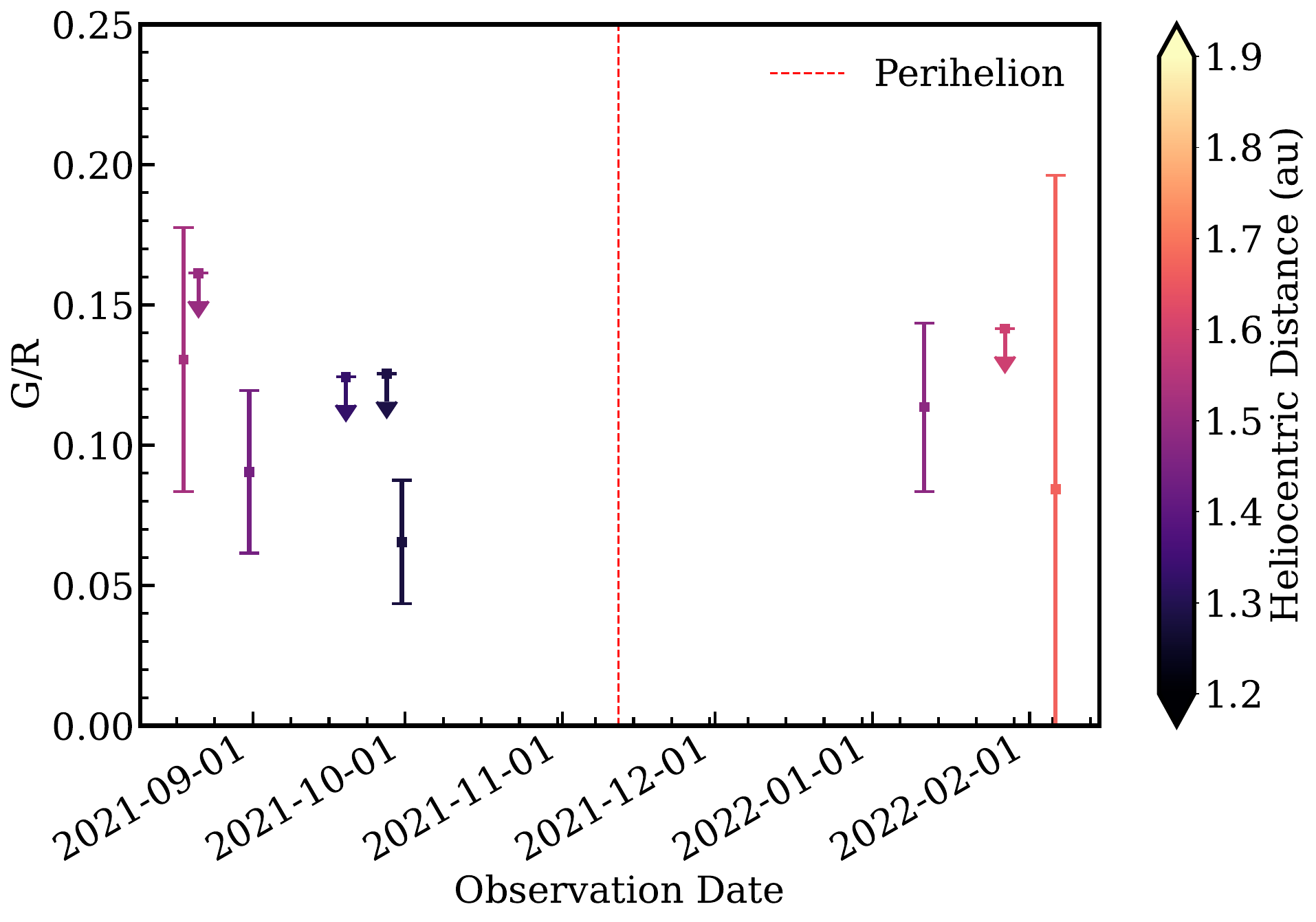}
        \caption{Evolution of the G/R forbidden oxygen ratio over heliocentric distance. Uncertainties are reported for observations with sufficient signal and observational conditions within constraints. Upper limits are reported for observations that do not satisfy signal or observing constraints.}
        \label{fig:GR}
    \end{figure}

Using the dust-subtracted 67P datacubes, we independently extracted summed spectra from a 5000 km radius cometocentric circular aperture and fit a single Gaussian to each of the 5577.339, 6300.304, and 6363.776~\AA~ [OI] lines. We chose a 5000 km aperture to allow for additional reference sky emission measurements to be taken from the corners of the observations, discussed in the next paragraph. We applied the same apertures and fits to the sky-observation datacubes, but also corrected for the flat sky continuum via a 1st-order polynomial subtraction in each region. We extracted fluxes from the fitted Gaussians, and summed the fluxes from the red-doublet lines. We can combine both red-doublet emissions into one channel since their intensities arise from the transition between the (2p$^{4}$)$^{1}$D$_{2}$ state to the (2p$^{4}$)$^{3}$P$_{1,2}$ ground states. \cite{Decock:2013} noted the red-doublet intensity ratio should be synonymous with the 6300.304 and 6363.776~\AA~ branching ratio, which is $\sim$3 as determined by \cite{Galavis:1997} and \citet{Storey:2000} via quantum mechanical emission models. We leveraged this known ratio to check for possible NH$_{2}$ contamination in our red-doublet flux, which would have artificially enhanced the red-doublet ratio to much greater than 3. We found significant contamination across several nights, which limit our sensitivity and prohibit characterisation beyond the presentation of upper limits (see Figure \ref{fig:GR}). Despite this contamination, the [OI] flux still dominated the NH$_{2}$ emissions and allowed for measurements across the campaign. 

Following the extraction of aperture spectra and flux measurements, we applied a correction for atmospheric [OI] by subtracting the green and red-doublet sky emissions from the 67P [OI] measurements. Initially, we attempted a direct subtraction of sky emission from the 67P emission; however, this approach resulted in an over-subtraction of the 5577.339 ~\AA~ line, leading to negative values in most of the dataset. Fortunately, the 5577.339 ~\AA~ emission from 67P was confined to the central few thousand kilometres of the comet's coma, leaving large portions of the datacube unaffected by cometary emission. These comet-free regions provided an alternative sky-observation for the green [OI] line, which significantly improved the accuracy of the sky correction. We did not encounter similar problems with the sky correction of the 6300.304 and 6363.776 ~\AA~ [OI] lines, so we used the measured fluxes from the dedicated sky-observations.

After we propagated the atmospheric [OI] corrections, we computed the green to red-doublet ratio for 67P, as seen in Figure \ref{fig:GR}. We estimated uncertainties from the original Gaussian fits. Larger uncertainties are observed at greater heliocentric distances due to the non-proportional subsidence of [OI] emissions versus contaminating NH$_{2}$ emissions, where NH$_{2}$ remains stronger for longer, therefore increasing contamination and uncertainty. The last G/R was computed around 1.7 au on 06 February, and constituted the furthest heliocentric distance probed. Naturally, this measurement has the largest uncertainty and likely occupies the upper range of the uncertainty interval, rather than a G/R $\sim$0.1. Despite these constraints, we retrieve several measurements around 0.1 pre-perihelion (0.13$\pm$0.05 to 0.07$\pm$0.02) and post-perihelion (0.12$\pm$0.04). These are all consistent with a water-dominated coma. 

\subsubsection{Radial Haser Models} \label{subsubsec: haser}

Radial Haser models provide a framework for investigating the presence of extended sources in the coma. By fitting radial intensity profiles, it is possible to infer the contributions of primary sources (direct parent species sublimation from the nucleus) and extended sources (icy particles or dust) through comparing the fitted parent species scale lengths with the previously derived parent scale lengths in \cite{Fink:1991, Ahearn:1995, Bannister:2020}. These extended sources can release gas over larger distances, altering the apparent morphology of the coma and impacting the size of the parent scale lengths. 

    \begin{figure}[t!]
    	\includegraphics[width=\textwidth]{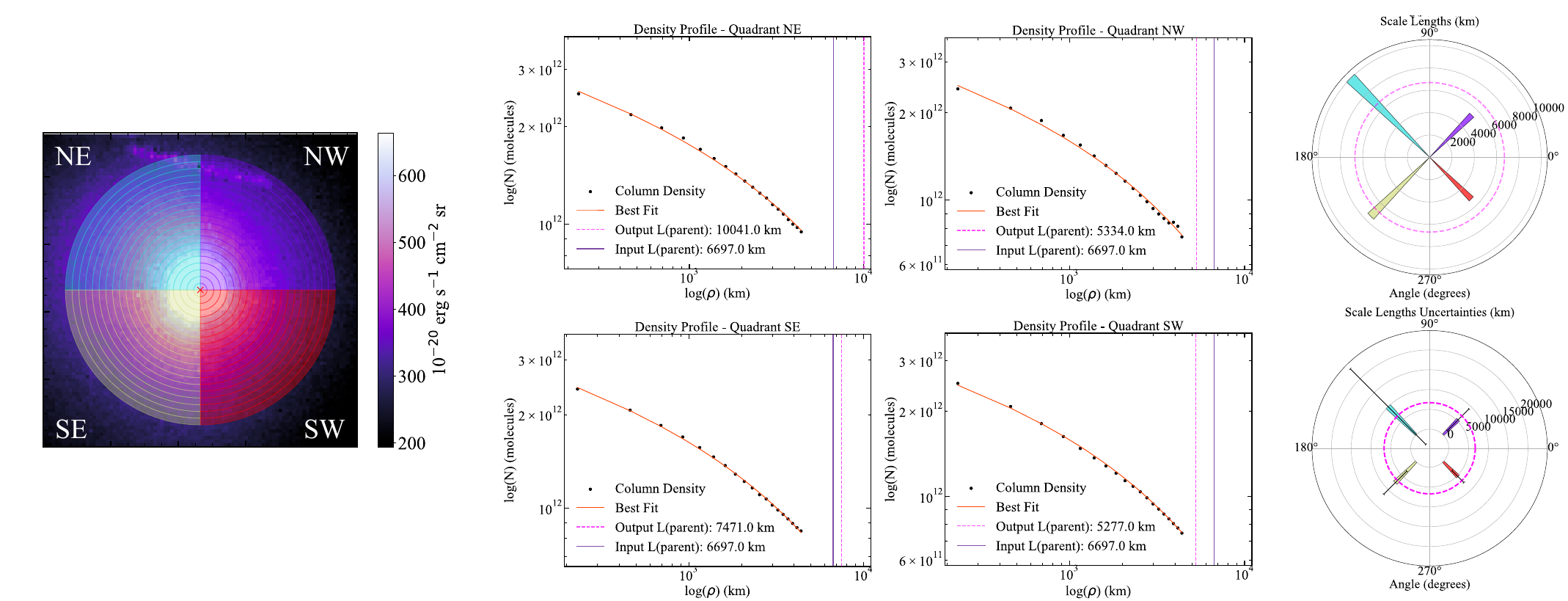}
        \caption{Radial Haser fits of the highest SNR NH$_{2}$ map, 30 September, 2021. Left-most panel depicts the unenhanced map split into quadrants and radial wedges, from which the density profiles in the middle panels were constructed from. The middle panels are the Haser fits (red line), profiles (black markers), input parent scale length (purple vertical line), and output parent scale length (pink vertical line). Finally, the right-most panels depict the different quadrant-specific output scale lengths (wedges) vs the input scale length (pink dotted circle), with associated uncertainties in the lower right-most panel.}
        \label{fig:haser}
    \end{figure}

    \begin{landscape}
    \begin{table}
        \caption{\label{tab:nh2_scales} Fitted Effective Scale Lengths for Pre-perihelion NH$_{2}$.}
        \centering
        \begin{tabular}{lcccc}
        \hline\hline
        Date & Input Scale Length & Fitted Avg Scale Length & Fitted NE Scale Length & Factor \\
        \hline
        2021-September-30 & 6698 & 6035 & 10025 & 1.7  \\
        2021-September-19 & 7225 & 6503 & 9598 & 1.5  \\
        2021-August-18 & 9498 & 11088 & 20888 & 1.9  \\
        \hline
        \end{tabular}
        \newline
        \newline
        \textbf{Notes. }{Parameters for the radial Haser profile fitting. Date is the observation epoch, and Input Scale Length (in kilometres) is the heliocentrically scaled starting value for the fit. The Fitted Avg Scale Length (in kilometres) is the average of the NW, SE, and SW quadrant fitted scale lengths, while the Fitted NE Scale Length (in kilometres) is solely from the NE quadrant representative of the known NH$_{2}$ spiral. Factor is the ratio of Fitted NE Scale Length to Fitted Avg Scale Length.}
    \end{table}
    \end{landscape}

We apply radial Haser profile fitting to dissect the extended production mechanisms of CN and NH$_{2}$, using methods most recently described in \cite{Ferellec:2024}. We identified potential contributions from extended sources for NH$_{2}$, however, maps in other species had insufficient SNR to analyse in this way. 

First described in \cite{Opitom:2019} for MUSE data, we constructed four NH$_{2}$ radial profiles through wedge-based aperture photometry across different quadrants, seen in the leftmost panel of Figure \ref{fig:haser}. We converted the NH$_{2}$ flux to radial column density via Equation \ref{eqn:2}, in which $F_\mathrm{coma}$ is the radial flux, $g$ is the NH$_{2}$ fluorescence efficiency from \cite{Ahearn:1995}, and $\Delta$ is the geocentric distance in au. 

\begin{equation} \label{eqn:2}
     N_\mathrm{coma}(\rho) = F_\mathrm{coma}(\rho) * \frac{1}{g} * (4 \pi \Delta^{2}) 
\end{equation}

We then utilise \texttt{SBPY} \citep{Mommert:2019} to implement the classic \cite{Haser:1957} model, described in equation \ref{eqn:3}, where $Q_{0}$ is an \textit{a priori} estimate of the NH$_{2}$ production rate, $\rho$ is cometocentric distance, $v_\mathrm{gas}$ is gas expansion velocity, and $L_\mathrm{p}$ and $L_\mathrm{d}$ are known parent and daughter species scale lengths.

\begin{equation} \label{eqn:3}
     N_\mathrm{haser}(\rho) = \frac{Q_{0}}{(4 \pi \rho^{2}v_\mathrm{gas})} \frac{L_\mathrm{d}}{L_\mathrm{p}-L_\mathrm{d}} ( \exp{(-\frac{\rho}{L_\mathrm{p}})} - \exp{(-\frac{\rho}{L_\mathrm{d}})})
\end{equation}

\noindent We take values of $v_\mathrm{gas}$, $L_\mathrm{p}$, and $L_\mathrm{d}$ previously tested on MUSE data from Table 2 in \cite{Bannister:2020}, and NH$_{2}$ $Q_{0}$ estimates during the 2009 67P apparition from Table 3 in \cite{Lara:2011}. We scaled $L_\mathrm{p}$ and $L_\mathrm{d}$ values with heliocentric distances. We utilise least squares fitting, varying $L_\mathrm{p}$ and $Q_{0}$, to fit the Haser model ($N_\mathrm{haser}(\rho)$) to the observational data ($N_\mathrm{coma}(\rho)$). We cannot vary $L_\mathrm{d}$ due to the proximity of 67P, which limited the FoV to well within expected daughter species scale lengths (10$^{4}$-10$^{5}$ km). 

We attempted this process on all molecular maps, across all epochs, however, only the brightest NH$_{2}$ maps bounding perihelion yielded satisfactory fits. Depicted across the middle four panels of Figure \ref{fig:haser}, we achieved high-fidelity fits for each of the quadrants. Intriguingly, the fitted parent scale length is enhanced by a factor of 1.7 from the rest of the scale lengths in the coma, specifically in the NE quadrant that encompassed the large, curved, and stable NH$_{2}$ pre-perihelion structure. Similarly, the profile for the NE quadrant was visually extended beyond those of the other quadrants. We computed uncertainties for the fitted parameters through the least square covariance matrix, and plotted them in the bottom right panel of Figure \ref{fig:haser}. While the uncertainties are large, it is expected that the quadrant with the largest parent scale length deviation would also have the largest uncertainty. If we were able to vary the daughter species scale length, we would achieve significantly lower uncertainties for our fits. Therefore, we can only report that parent scale lengths in the NE quadrant of 30 September 2021 are consistent with possible extended source contribution to the observed NH$_{2}$ morphology. Notably, we find similar visual and scale length elongations in the NE quadrant across the pre-perihelion epochs (1.5$\times$ on 19 September, with the largest parent scale length ratio of 1.9 on 18 August, see Table \ref{tab:nh2_scales}). This suggests that the sustained pre-perihelion emission from this structure could be consistent with either partial or full emission from extended sources. 

\section{Discussion} \label{sec:discussion}

\begin{figure}[hbt!]
        \centering
    	\includegraphics[scale=0.55]{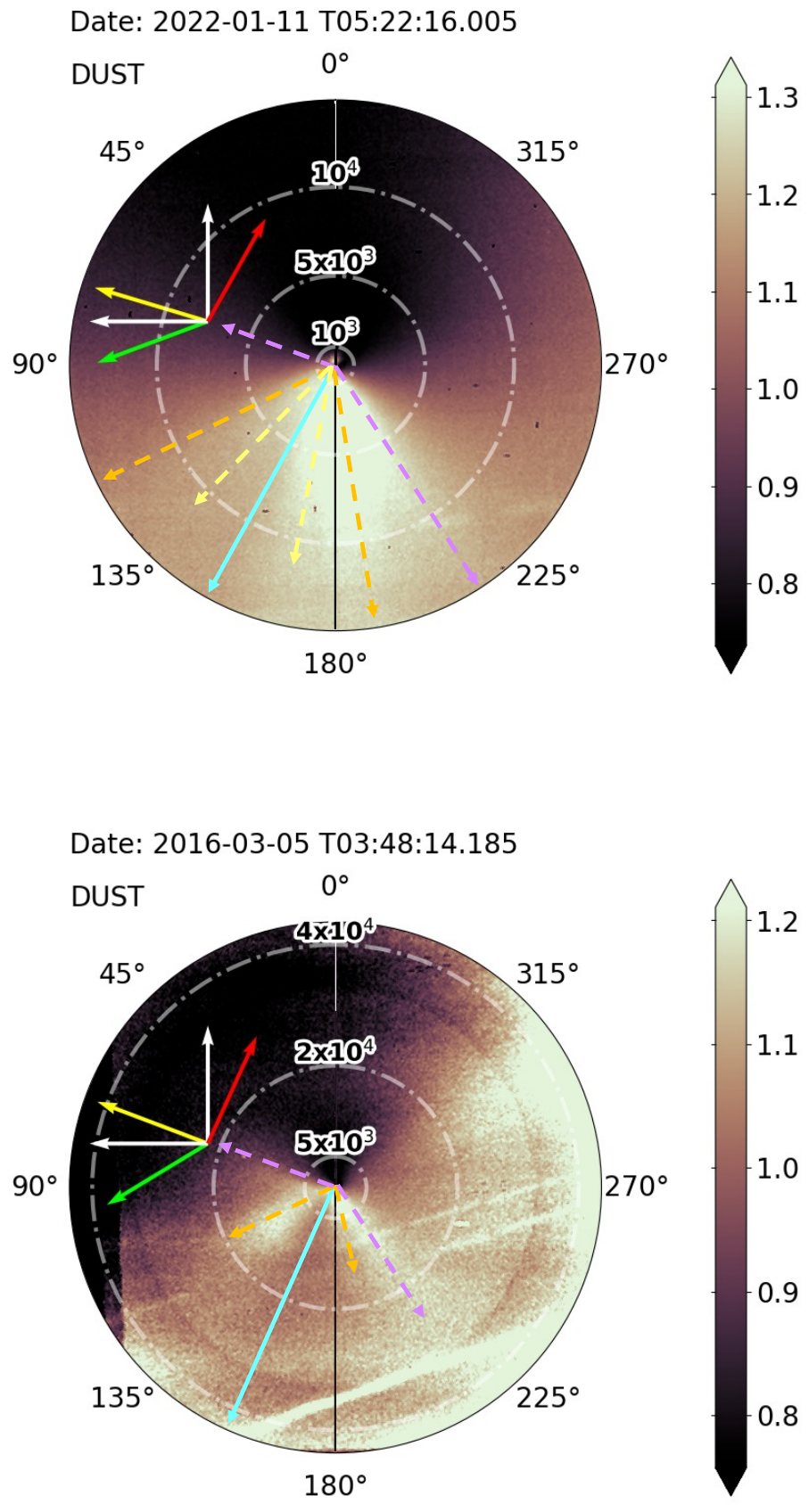}
        \caption{Comparison of post-perihelion fans from 2022 (top panel) and 2016 (bottom panel). The red, lime, yellow, and white arrows are the same annotations as Figures \ref{fig:dust_enh} through \ref{fig:cn_enh}. The concentric rings denote cometocentric distance, in units of kilometres. The purple dashed line corresponds to active region between -10$^{\circ}$ to -30$^{\circ}$ S. The orange dashed line corresponds to -40$^{\circ}$ to -50$^{\circ}$ S. The yellow dashed line corresponds to -60$^{\circ}$ to -80$^{\circ}$ S. The cyan line is the South Pole of 67P. Binning is 1$\times$1.}
        \label{fig:compare}
    \end{figure}

Ground-based observations provide critical context for understanding the chemical and physical evolution of the coma. Recent radio, infrared, and optical studies by \cite{Biver:2023}, \cite{Bonev:2023}, \citet{Boehnhardt:2024}, and \citet{Ivanova:2024} examined the dust and gas environment of 67P during its 2021 apparition, both pre- and post-perihelion. Notably, the studies identified dust structures that align closely with the temporal and angular ranges of multiple structures and fans observed in our dataset. \cite{Boehnhardt:2024} utilised Laplacian filtering techniques to detect several fans, neck-line features, and streamer structures across 33 epochs spanning 21 May 2021 to 1 June 2022, which best complements our dataset. Their work revealed four dynamically modelled dust fans located at latitudes of +40$^{\circ}$ N and –10$^{\circ}$ S pre-perihelion, and –10$^{\circ}$ S, –50$^{\circ}$ S, and –70$^{\circ}$ S post-perihelion. The post-perihelion fans at –10$^{\circ}$ S, –50$^{\circ}$ S, and –70$^{\circ}$ S correspond to active regions previously identified by \cite{Boehnhardt:2016} and recently modelled using ROSINA measurements by \cite{Lauter:2022}. Further works from the 2015 apparition also report dust structures similar to our own \citep{Knight:2017}. Our Fans B through F show a coarse correlation with their reported –10$^{\circ}$ S, –50$^{\circ}$ S, and –70$^{\circ}$ S active regions. It should be noted that pre-perihelion Fan B is likely just post-perihelion Fans C through F superimposed onto each other due to the side-on viewing geometry pre-perihelion. \citet{Biver:2023}, \citet{Bonev:2023}, and \citet{Ivanova:2024} report finding similar dust fans across the apparition, and corroborate these findings with an active region around –58$^{\circ}$$\pm$5$^{\circ}$ S. We find similar structures in 2016 MUSE observations, seen in Figure \ref{fig:compare}, that further supports long-term stability of these dust features and activity sources over multiple apparitions. Lower latitude emission sources and structures were present (purple and orange lines, -10$^{\circ}$ to -30$^{\circ}$ S, -40$^{\circ}$ to -50$^{\circ}$ S), however, the highest latitude structures are not seen. This is most likely due to the 4$\times$ larger FoV in the 2016 MUSE data, and significantly higher heliocentric and geocentric distances of 2.5 and 1.5 au, respectively. In situ measurements by MIRO in 2015 \citep{Biver:2019} detected elongations of various parent molecules (NH$_3$, CH$_3$OH, CO) and water isotopologues (H$_2^{16}$O, H$_2^{17}$O, H$_2^{18}$O) originating from $\phi$ = -80$^{\circ}$$\pm$5$^{\circ}$ S, with a fan width of 100$^{\circ}$–130$^{\circ}$ around perihelion. This provides further evidence for the stability of active regions across multiple apparitions, tenuously links the Rosetta and Earth-based observational scales, and aligns well with the lowest latitude reported by \cite{Boehnhardt:2024} and our study. These findings also agree with our reported widths, position angles, and proposed southern activity sources for dust Fans C-F, and molecular structures of C$_2$, NH$_2$, [OI], and CN, as detailed in Subsections \ref{subsubsec: dust morph} and \ref{subsubsec: gas morph}.

    \clearpage    

One advantage of Laplacian filtering in \cite{Boehnhardt:2024} is the detection of very dim and sharp structures in the coma, generally beyond what division by azimuthal median can achieve. Such sensitivity allowed the detection and correlation of related fans. Instrumental artefacts, such as the cross-hatching pattern seen in several maps, prohibit us from using Laplacian filtering on our dataset. These related fans are associated with continuous ejection of material from an active source on the nucleus as it rotates, producing a conical distribution of material around the projected rotational axis. The detected fan features in \cite{Biver:2023, Boehnhardt:2024, Ivanova:2024} all correspond to the sides of this ejection cone, and therefore should be mirrored about the sky-projected rotation axis. Although \cite{Boehnhardt:2024} and \cite{Ivanova:2024} find evidence of mirrored fans pre-perihelion, we do not. We do find strong evidence post-perihelion for this behaviour between Fans C and E, as they are approximately equidistant from the South Pole of 67P. Therefore, we posit that Fans C and E originate from the same active source on the nucleus, which is likely the source $\phi$=-50$^{\circ}$ to $\phi$=-58$^{\circ}$ S reported in the aforementioned works. Fans D and F likely represent one component of a mirrored fan, however, we were not able to retrieve their complementary structures robustly in our CWT maps or Voigt fits, therefore limiting the speculation to their origin and sublimation latitude. Despite this, Fan D is closer to the South Pole of 67P, thus implying a higher sublimation latitude, likely on par with –70$^{\circ}$ S to –80$^{\circ}$ S, reported in \cite{Biver:2019} and \cite{Boehnhardt:2024}. Fan F is further from the South Pole, likely correlating to the -10$^{\circ}$ to -30$^{\circ}$ S sublimation zones found in both their and our work. 

We do not observe evidence of mirrored fans in our gas maps, however, we note a strong correlation between the pre-perihelion northern NH$_{2}$ structure and Fan A. We also find a significant post-perihelion relationship between all gas species and dust fans, though disentangling the post-perihelion interactions remains challenging. \cite{Boehnhardt:2024} similarly reported a pre-perihelion dust structure with clockwise curvature originating from northern regions of 67P, identifying a source at $\phi$ = 40$^{\circ}$ N. This latitude is approximately 40$^{\circ}$ south of the active region proposed in our study, a discrepancy we attribute to our limited sensitivity to mirrored fans, which reduced the precision of latitude conversions. Despite this limitation, our Fan A and NH$_{2}$ structure align well with the eastern component of the mirrored fan described in \cite{Boehnhardt:2024}, possibly indicating a shared activity source for dust particles and NH$_{2}$ around $\phi$ = 40$^{\circ}$ N. Further correlations can be drawn to in-situ findings by \cite{Biver:2019}, which identified the majority of gas sublimation originated from high northern latitudes during the pre-perihelion period (July 2014 to February 2015), including contribution from parent species NH$_{3}$. In-situ pre-perihelion results from \cite{Lauter:2022} show prolonged northern emission of NH$_{3}$ between 0-40$^{\circ}$ N latitudes, consistent with this work and previous studies. Finally, the global distribution of NH$_{3}$ activity by mass fraction is nearly uniform (around 0.3$\%$) \cite{Lauter:2022}. This implies that NH$_{2}$ signal should be relatively uniform from north to south, however, we observe the opposite in our maps where the NE quadrant is 10-15$\%$ more intense than the rest of the coma, suggesting additional sequestered sources of NH$_{2}$ in the north.  While photochemical scale lengths for this region are enhanced by 1.5-1.9$\times$ the average fitted values for the rest of the coma, it is most likely the northern pre-perihelion NH$_{2}$ signal is coming from a mix of both nucleic NH$_{3}$ sublimation and possible extended source fragmentation. 

One possible interpretation for an extended source of NH$_{2}$ could be ammonium salts (NH$_{4}^{+}$X$^{-}$), including ammonium formate and ammonium sulfate in 67P’s highly porous refractory dust, initially identified by VIRTIS \citep{Poch:2020} and ROSINA instruments \citep{Altwegg:2020}. Evidence from Rosetta’s COSIMA instrument indicates that 67P's dust is composed of nearly equal amounts of high molecular weight organics and anhydrous mineral phases (45-55$\%$ wt), with up to 90$\%$ porosity \citep{Fray:2016, Langevin:2017}, which could provide a mechanism to preserve NH$_{3}^{+}$X$^{-}$ further from the nucleus. \cite{Güttler:2019} revealed that solid, fluffy, and porous dust particles undergo complex fragmentation and fading processes as they move outward from the nucleus, ultimately depositing any hidden volatiles into the coma. In this context, we argue that the northern pre-perihelion NH$_{2}$ signal could partially result from fragmentation and sublimation of porous, organic-rich dust particles that could release semi-volatile species like ammoniated salts along their trajectories, which further dissociate into NH$_{2}$ upon exposure to solar radiation \citep{Altwegg:2020, Poch:2020}. 

One trend in our maps is the detection of CN signal solely from the southern hemisphere of 67P. \cite{Opitom:2017} first reported the emergence of southern CN signal after the vernal equinox, likely caused by the change of season and migration of the subsolar point to the southern hemisphere of 67P. Similarly, observations by \cite{Knight:2017} from 2015 report broad CN fans from the southern hemisphere, which further confirmed this trend. This change in seasons illuminated long-dormant (5.5 yr) reservoirs in the South, which intensely sublimated and both exposed new volatile-rich substrate and lofted larger dust particles. Intriguingly, this clear dependence was not detected from in-situ ROSINA measurements, which suggests the presence of an extended CN source that breaks down at distances further than several hundred kilometres from 67P. However, VIRTIS and ROSINA measurements highlight that the hemispherical difference in CN is not a product of 67P's formation, but is rather a result of de-volatilisation across successive 5.5 yr northern summers \citep{Bockelée-Morvan:2016, Fougere:2016}. 

We observe this strong seasonal dependence and southern enrichment of CN post-equinox (6 August 2021), and find evidence for a correlation between the dust Fan B and a highly collimated CN jet (both originating from southern mid-latitudes). While we do eliminate the possibility of dust contamination in our CN maps (see Appendix \ref{appendix: gas-dust}), we were unable to retrieve satisfactory fits for parent species scale lengths, thus limiting our interpretation of possible extended CN sources. Nonetheless, the strong correlation of dust Fan B and the pre-perihelion CN jet provides tentative evidence for the existence of CN distributed sources. This interpretation agrees with previous studies on possible extended sources of CN, e.g. in 1P/Halley \citep{Ahearn:1986}. We do see an enhancement in particle colour, and thus larger size or unique compositions, in the same region as the CN structure and dust Fan B from 31 August to 30 September. These larger red dust particles could similarly act as a transport mechanism for CN parent species, as we proposed for NH$_{2}$ parent species. A possible scenario is that larger disintegrating dust particles, perhaps containing CHON particles or macromolecules within their porous interiors, could be a transport and source mechanism for the observed CN jet, rather than simply CHON particles or HCN nucleus outgassing alone. It should be noted that other parent molecules have been identified sublimating from southern mid-latitudes, such as HCN \citep{Lauter:2022}, which suggests both extended sources and traditional nucleic sublimation as sources for this CN structure and CN in the southern hemisphere.

Extensive studies of nucleus and coma colour were conducted throughout the Rosetta campaign \citep{Fornasier:2015,Ciarniello:2016,Filacchione:2016,Fornasier:2016}, in which a strong orbital and seasonal dependences were found. These studies target different wavelength ranges from $\sim$4000 to 10000~\AA, ground-based and in-situ. Nonetheless, the general trends are best described in \cite{Filacchione:2020}, across two windows from 4000 to 5000~\AA~and 5000 to 8000~\AA. Generally, pre-perihelion, the nucleus exhibits a reddish colouration, dominated by carbon-rich particles and complex organics, while the coma is primarily populated by blue, ice-rich dust particles. As activity increases, more ice-rich particles fill the coma, creating an increasingly blue coma. As the comet nears perihelion, increased activity lifts the redder surface layers off the nucleus and into the coma, gradually reddening the coma. By perihelion, the nucleus is significantly bluer than the coma, which should be at its reddest. During the outbound trajectory, large red dust particles in the coma are replaced by smaller, bluer material lifted from the now-blue nucleus, transitioning the coma from red to blue. While our maps cannot probe the coma during the reddest perihelion stage, we do note similar pre-perihelion and post-perihelion bluening trends reported by the aforementioned studies. While phase-reddening is shown to cause significant colour deviations at high angles ($>$ 50$^{\circ}$) \citep{Bockelée-Morvan:2019}, it has negligible effects on our work (generally low phase angles, except 27 and 30 September).  

Further 2014-2016 in-situ and Earth-based studies report spectral slopes at comparable optical wavelengths, from 11-16$\%$/1000~\AA around perihelion, which are consistent with our results (8.2-18.4$\%$/1000~\AA) \citep{Fornasier:2015, Bertini:2017, Opitom:2020}. We can further compare our spectral slopes with ground-based broadband colour determinations by \cite{Gardener:2022}, where their g$-$r and r$-$i measurements most closely represent our spectral slopes from 5000 to 7000~\AA. Therein, \cite{Gardener:2022} reports broadband colour measurements from the 2021 apparition which are consistent with a stable coma colour throughout perihelion ($\sim$0.6 colour index g$-$r, $\sim$0.2 colour index r$-$i). However, these measurements have significant scatter ($\pm0.2$), which would hide any subtle trends around perihelion, like our gradual bluening (a few to several $\%$/1000~\AA). Additional recent studies from the 2021 apparition report higher spectral slopes of 15.2$\pm$4.1$\%$/1000~\AA~(6 October 2021) in the spectral range 4450–6839 ~\AA, and 14.2$\pm$0.4$\%$/1000~\AA~(6 February 2022) in the same range \citep{Ivanova:2024}. Two of our measurements coincide with these reports, notably 30 September at 11.2$\%$/1000~\AA~and 06 February at 8.3$\%$/1000~\AA. While we do not report uncertainties with our maps, an uncertainty of $\pm$1-2$\%$/1000~\AA~should be assumed in all of our slope measurements. These uncertainties are estimated by varying the spectral window we extract slopes from, and calculating the mean difference of the different fitted regions. Furthermore, the discrepancy is also due in part to the different spectral ranges (5000-7000 ~\AA), which are known to cause a steeper slope at shorter wavelengths. The same reasoning cannot justify the nearly 5$\%$/1000~\AA~discrepancy on 6 February on its own, and is likely due to the additional combination of differing apertures, use of broadband filters (including gas contamination), differing reference solar spectra, and instrument-specific spectral calibration techniques.

The seasonal evolution of H$_{2}$O and CO$_{2}$ activity around 67P is highly dynamic and influenced by heliocentric distance and hemispheric insolation. We find evidence for hemispherical H$_{2}$O anisotropy in our red-doublet [OI] maps, where H$_{2}$O emission dominated the flux and distinct substructures are detected (Fig \ref{fig:OI_enh}). We attempted to find similar anisotropic substructure in green [OI] maps (which is populated by significant CO$_{2}$ and CO emission), however, none were conclusively detected, which could be due to the large SNR difference between the green and red-doublet maps. During perihelion, the comet's southern hemisphere, which is in summer, undergoes significant erosion of a few meters, exposing volatile-rich subsurface layers \citep{Keller:2015}. This process contributes to a water-dominated coma, as evidenced by our consistent G/R measurements around or less than 0.1. Ideally, this ratio would change with heliocentric distance, however, we are limited by the spectral resolution and sensitivity of MUSE, the suboptimal observing conditions, and the possible contamination of NH$_{2}$. Therefore, can only report that our G/R trend is consistent with a coma dominated by H$_{2}$O across our observational campaign. 

Finally, we find a strong morphological and evolutionary correlation between the [OI] and C$_{2}$ maps, giving possible insights into the parent species and release mechanism of C$_{2}$. It has been shown that the majority of water sublimation on 67P is related to the location of the subsolar point, or point of maximum insolation on the nucleus \citep{Keller:2015, Lauter:2022}. We find that the approximate position angles of the [OI] structures in Figure \ref{fig:OI_enh} roughly correlates to the position of the subsolar point on the nucleus, as well as the C$_{2}$ structures (Fig. \ref{fig:c2_enh}). Thermophysical models, based on ROSINA measurements, depict similar evolutionary trends and sublimation zones between H$_{2}$O and C$_{2}$H$_{6}$, especially in the southern hemisphere \citep{Lauter:2022}. These identified active spots directly sublimate H$_{2}$O and C$_{2}$H$_{6}$, along with many other trace species \citep{Longobardo:2019}. We do not find evidence for extended sources of C$_{2}$ or [OI], confirming this trend, and pointing toward direct C$_{2}$ photodissociation from hydrocarbon parent species like C$_{2}$H$_{6}$, among others.

\section{Conclusions}

In this work, we utilised MUSE observations of comet 67P/Churyumov-Gerasimenko to generate simultaneous maps of dust, C$_2$, NH$_2$, [OI], and CN comae across 12 observational epochs pre- and post-perihelion. These maps were employed to investigate the evolution and composition of 67P's middle and outer coma during its most recent apparition, placing our findings within the broader context of Rosetta's 2014-2016 results from the nucleus and inner coma environments. We did not carry out detailed modelling, as it is outside the scope of this paper. Our major findings are:

\begin{enumerate}
 \item We deconstructed the gas-dust coma into constituent and concurrent maps of dust and gas species. We followed the morphological evolution of dust and gas species across the apparition, and found well-known dust fans and structures \citep{Boehnhardt:2016,Biver:2023,Bonev:2023,Boehnhardt:2024,Ivanova:2024}. We also found gas signal from C$_2$, NH$_2$, [OI], and CN, and noted the unique association of NH$_2$ signal and CN signal with the aforementioned dust structures. Finally, our results build upon the accepted notion of long-term stability of active sources on 67P, across multiple apparitions.
 \item In the gas maps, we identified two pre-perihelion evolutionary regimes: (1) evolving H$_2$O ([OI]$^{1}$D) and C$_2$ signal, (2) stable NH$_2$, dust, and CN signal. We present arguments, in the context of other Earth-based and in-situ Rosetta studies, that different production processes are responsible for the two regimes. We propose that the (1) evolving regime is linked to sublimation directly from the nucleus and subsolar evolution (both molecules), while the (2) stable regimes are the joint product of distributed coma sources (NH$_2$, CN) and seasonal effects. 
 \item We investigated the connection of NH$_2$ and CN to known dust structures, and probed their relation to emission from extended sources. We also find that the pre-perihelion northern NH$_2$ structure has photochemical scale lengths 1.5-1.9$\times$ the rest of the coma, which suggests a mechanism that preserves the parent species until further from the nucleus, thus enhancing the effective scale length. We propose a possible extended source of ammoniated salts \citep{Altwegg:2020,Poch:2020}. Signal-to-noise restricted this similar analysis of CN and other species, however, we performed tests to ensure that the CN signal corresponding to Fan B was not contaminated by residual dust flux. We did not find any dust flux, and confirmed the CN signal to be real. However, we do see larger, redder dust particles in our colour maps which correspond to both the CN and dust Fan B morphologies, which could explain why CN signal is found with the dust since parent species would be preserved in these larger dust particles for longer. Finally, in the context of Rosetta results which show an excess in CN signal that cannot be explained by HCN photolysis alone directly from the nucleus, we propose that CN is also produced in-part by extended sources in the south \citep[e.g.,][]{Hanni:2020}. However, we do note that NH$_2$ and CN signal is likely to come from both traditional nucleus sublimation and these distributed sources \citep{Lauter:2022}.  
 \item We computed dust spectral slope maps, in which the spectral slope, or colour, of the dust could be quantified. Largely, we report evolutionary trends consistent with Rosetta findings \citep{Filacchione:2020}.
 \item We calculated the G/R [OI] ratio across perihelion to probe the H$_2$O to CO$_2$/CO dominance of the coma. We found G/R ratios were generally consistent with an average value of 0.1. This implies that the coma is dominated by water sublimation throughout our observational campaign. 
\end{enumerate}

\section*{Acknowledgements}

For the purpose of open access, the author has applied a Creative Commons Attribution (CC BY) licence to any Author Accepted Manuscript version arising from this submission. This analysis was based on observations collected at the European Southern Observatory under ESO programmes 105.2086.001 and 108.223B.001, and archival ESO programmes 096.C-0160(A) and 096.C-0855(B) taken by PI Aurélie Guilbert-Lepoutre. This work was supported by the International Space Science Institute (ISSI) in Bern, through ISSI International Team project 504 “The Life Cycle of Comets”. The authors similarly wish to acknowledge the dedication and technical support of the support astronomers, Instrument Operators, and Telescope Operators at Paranal Observatory. 

For the purpose of open access, the authors have applied a Creative Commons Attribution (CC BY) licence to any Author Accepted Manuscript version arising from this submission. 

\section*{Data Availability}

This commitment to data accessibility is in line with the principles of transparency and reproducibility, our research team, and the scientific community, facilitating the validation of our results and encouraging further scientific inquiry. All raw observations can be collected at the European Southern Observatory Archive under the aforementioned ESO programme IDs. Processed data is available upon request. 

\section*{Declaration of generative AI and AI-assisted technologies in the writing process.}
During the preparation of this work the author(s) used OpenAI services to increase clarity of syntax and grammar. After using this tool/service, the author(s) reviewed and edited the content as needed and take(s) full responsibility for the content of the publication.

\appendix

\section{Molecfit Modelling Parameters} \label{appendix: molecfit}

In order to effectively model the transmission spectra using Molecfit, we identified regions of atmospheric gas emission and target gas emission, shown in Table \ref{tab:molecfit_params}. The included wavelengths are regions in which atmospheric signature needs to be modelled and taken into account, while the excluded wavelengths are regions populated by non-atmospheric emission that the fitting software should not include. The excluded wavelengths are always regions within the larger included atmospheric windows. We used the standard atmospheric windows provided by the European Southern Observatory. 

    \clearpage    
    \begin{table}
        \caption{\label{tab:molecfit_params} Atmospheric Wavelengths Modelled by Molecfit}
        \centering
        \begin{tabular}{cc}
        \hline\hline
        Included Wavelengths (~\AA) & Excluded Wavelengths (~\AA) \\
        \hline
        5876.5-5998.6 & 5976.2-5995.8\\
        6453.5-6604.9 & 6534.3-6548.7\\
        6864.6-7076.4 & 6918.7-6948.4\\
        $''$ & 6956.9-6977.0\\
        $''$  & 6994.0-7039.0\\
        7160.1-7414.7 & 7244.1-7248.5\\
        $''$ & 7342.0-7354.3\\
        $''$ & 7367.5-7381.1\\
        7676.5-7711.2 & 7681.8-7701.9\\
        8133.4-8341.4 & none\\
        8945.2-9199.3 & 9138.3-9274.1\\
        \hline
        \end{tabular}
        \newline
        \newline
        \textbf{Notes. }{The Molecfit wavelengths regimes that it should include vs that it shouldn't include. Exclusion is derived from gas emission (target not atmospheric) in the included wavelength ranges.}
    \end{table}

\section{Gas-Dust Disentanglement Test} \label{appendix: gas-dust}

\begin{center}
        \begin{figure}[htb!]
        \centering
    	\includegraphics[scale=0.48]{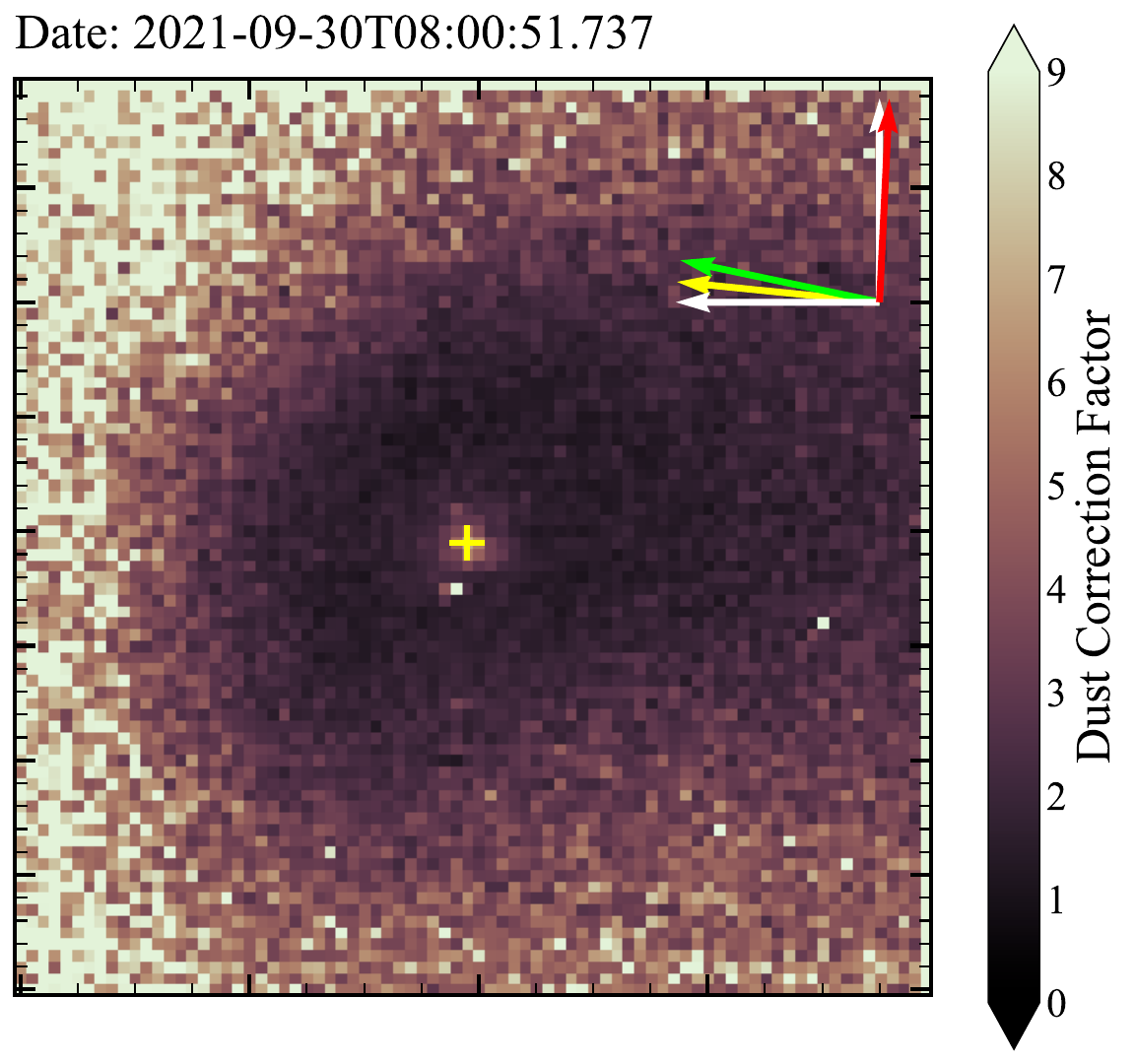}
        \caption{Normalised chi-squared map to investigate any residual dust-continuum in the CN map from 30 September 2021. Bright regions correspond to large continuum deviations and poor fits, while darker regions represent good fits. The yellow cross is the cometary optocenter. Yellow arrow is the sunward angle. Green arrow is the positive velocity vector of the comet. Red arrow is the sky-projected North Pole of 67P.}
        \label{fig:dust_check}
    \end{figure}
    \end{center}

As mentioned in Subsection \ref{subsubsec: gas morph}, we detect a strong correlation between dust Fan B and CN signal in the final pre-perihelion observation, 30 September 2021. We initially assumed dust contamination due to the near-perfect correlation between these features, so we sought to determine the spatial quality of the modelled dust-continuum we subtracted to make the maps. We implemented a chi-squared goodness of fit between the final shape- and slope-corrected reference dust spectrum with each input comet spectrum for the 30 September CN map. We created normalised chi-squared maps, shown in Figure \ref{fig:dust_check}, where the best subtractions are represented by the darkest values, while deviations (over and under subtractions) are denoted by higher dust correction factors. We had to normalise by the lowest chi-squared value due to the intrinsic deviation from chi-squared = 1 caused by the residual gas emissions across the spectrum. By implementing this normalisation with the lowest chi-squared, we then standardise the maps to reflect deviations from our best subtraction, therefore highlighting any residual dust contamination. 

    \clearpage    

We did not find any residual dust signal in the coma region corresponding to dust Fan B and the CN signal. To search for more diffuse substructure, we implemented division by azimuthal median enhancement to the normalised chi-squared map, and similarly found no residual dust signal in the region of Fan B in the CN map. Therefore, we maintain high confidence that the relation between dust Fan B and the CN structure is a product of in-situ emission and production processes, not from our data processing. 

\section{Continuous Wavelet Transform} \label{appendix: CWT}

We introduced the continuous wavelet transform in Subsection \ref{subsec:image processing}, which we used to help identify regions in the dust coma that corresponded to possible fan edges or mirrored fans. 

    \clearpage    
    \begin{figure}[hbt!]
        \begin{center}
    	\includegraphics[scale=0.69]{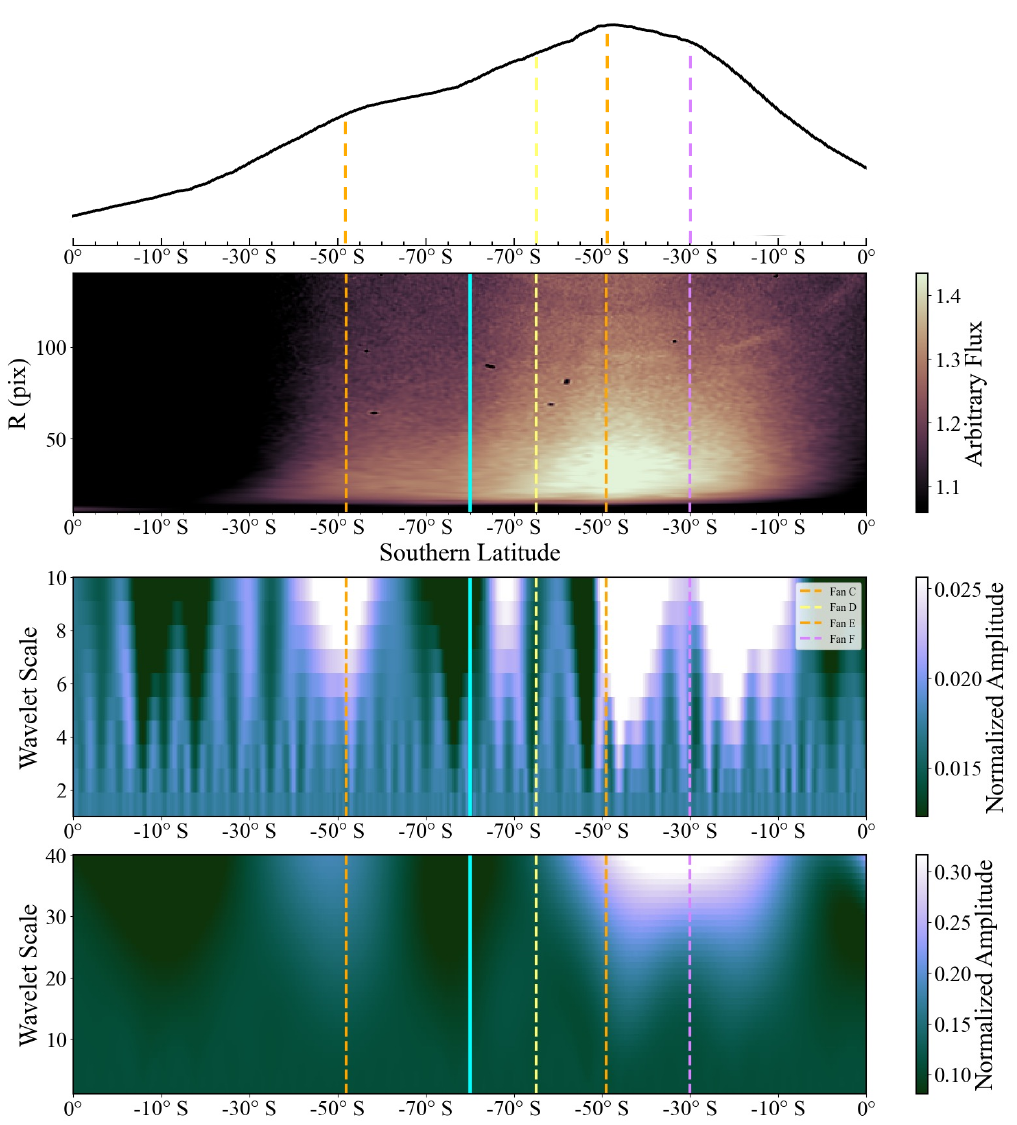}
        \end{center}
        \caption{Two-dimensional decomposition of flux profile on 11 January 2022 using two Continuous Wavelet Transform (CWT) regimes. The upper panel represents the extracted profile with Voigt fit solutions, the second panel depicts the enhanced Polar image Voigt fit solutions, the third panel represents small-scale CWT of the profile, which highlighted fine sharp substructure within the coma, and used wavelet widths from 1-10. The lowest panel is the larger-scale CWT, which therefore highlighted broad wide structures in the coma, and used widths from 1-40. Both x-axes correspond to southern latitudes, with the South Pole represented by the cyan. The y-axes on the lower two panels denote the wavelet scale (widths). The dashed lines denote angles measured by Voigt fits for Fans C-F. Fans C and E are the orange dashed line. Fan D is the yellow dashed line. Fan f is the purple dashed line.}
        \label{fig:CWT}
    \end{figure}

\bibliographystyle{elsarticle-harv} 
\bibliography{elsarticle-template-harv.bib}

\end{document}